\documentclass[10pt,aps,pra,final,twocolumn,amsmath,amssymb,longbibliography]{revtex4-1}

\usepackage[pdftex, colorlinks=true, linkcolor=blue, citecolor=blue, urlcolor=blue]{hyperref}
\usepackage{graphicx}
\usepackage{aurical}
\usepackage[T1]{fontenc}
\usepackage{units}

\newcommand{\ipcms}{Universit\'e de Strasbourg, CNRS, Institut de Physique et Chimie des Mat{\'e}riaux de Strasbourg, UMR 7504, F-67000 Strasbourg, France}

\begin{document}
	
\title{Deep neural networks for inverse problems in mesoscopic physics: Characterization of the disorder configuration from quantum transport properties}

\author{Ga\"etan J.\ Percebois}
\affiliation{\ipcms}
\author{Dietmar Weinmann}
\affiliation{\ipcms}

\begin{abstract}
We present a machine learning approach that allows to characterize the disorder potential of a two-dimensional electronic system from its quantum transport properties. Numerically simulated transport data for a large number of disorder configurations is used for the training of artificial neural networks. We show that the trained networks are able to recognize details of the disorder potential of an unknown sample from its transport properties, and that they can even reconstruct the complete potential landscape seen by the electrons.  
\end{abstract}

\maketitle

\section{Introduction}
\label{sec:introduction}

The precise understanding and the subsequent control of electronic transport through nanostructures is a research domain of great fundamental interest. It is also of crucial importance for the development of modern electronic devices.
While the ongoing miniaturization of electronic devices leads to an increasing importance of quantum effects, it simultaneously enhances the sensitivity of transport properties to material imperfections such as crystal defects or random dopant positions. These imperfections lead to an uncontrolled disorder configuration
that varies from one sample to another. As a consequence, the transport properties of samples fluctuate even if the macroscopic experimental control parameters are identical. Mesoscopic conductors are known for exhibiting sample-dependent universal conductance fluctuations in the low-temperature transport \cite{lee1985} and measurable consequences of the displacement of a single scattering center \cite{feng86,jura09a}. 

Acknowledging the importance of the sample-to-sample fluctuations, we here aim at a precise characterization of two-dimensional materials, focusing on the microscopic disorder configuration of a sample.
To this end, we investigate the dependence of spatially resolved quantum transport properties on the disorder configuration of the sample. Such data can be obtained with the help of the Scanning Gate Microscopy (SGM) technique \cite{topinka2000science,topinka2001nature,sellier2011review}, which measures the conductance as a function of the position of a local potential perturbation. It is possible to calculate the SGM response of two-dimensional electron gases in quantum materials like semiconductor heterostructures, graphene, or transition metal dichalcogenide (TMD) monolayers for a given disorder configuration
\cite{jalabert2010,gorini2013,steinacher2018}. The SGM response depends critically on (even weak) disorder in a characteristic way, with a disorder-dependent branching pattern that forms in the electron flow \cite{topinka2001nature,braem2018,fratus2019}. A method that allows to solve the inverse problem, that is to determine the disorder configuration from the SGM response or other transport data that is experimentally accessible, is highly desirable.

A possible way to solve an inverse problem is through a machine-learning approach \cite{deeplearning2016} using Artificial Neural Networks (ANN) \cite{mcculloch1943}. An ANN can produce an output information that depends on the input data in a highly complex way. In the process of supervised learning \cite{deeplearning2016}, the many parameters of the ANN are adjusted such that the output approaches the desired target on a set of training data. Choosing the appropriate architecture of the ANN, and a large enough set of training data, the trained ANN can be expected to produce an output for new input data that approximates the correct answer. These concepts allow to perform complex tasks with high precision.

The use of ANNs in physics is rapidly growing \cite{ML-RevModPhys2019}. Notwithstanding, the procedure is not based on a description of physical phenomena using model calculations. Even though a trained ANN might be powerful in predicting the behavior of a system, the details of the learning process and the mechanisms underlying the resulting predictions are difficult to extract. This approach is thus in sharp contrast to the usual model building that allows to understand physical phenomena and mechanisms. Interestingly, varying the shape of the ANN, it has been shown that the complexity of a possible theory to describe a set of experimental data can be determined \cite{ML-concepts2020}, and efforts are being made to better understand the functionings of trained ANNs \cite{voosen_XAI2017}.   

In this work we develop a machine-learning approach to characterize the disorder potential from electronic transport properties. To enable the ANN to recognize the details of the disorder potential, training with rich transport data for a large number of disorder configurations is needed.
To this end, the SGM response or similar electronic transport properties are computed for many microscopically different random disorder configurations of a two-dimensional electron system in a quantum material with a given sample geometry. Then, the resulting transport data, for which the underlying disorder configuration is known, are used to train appropriately shaped and parametrized ANNs. The trained ANNs are able to recognize various features of the disorder configuration from the input of transport data. The performance of the trained ANNs is characterized using calculated data that have not been used in the training process.
Although we focus on electronic transport in two-dimensional electron gases, the approach is more general and can be applied to various situations, as for example the propagation of light in films of varying thickness \cite{patsyk20} or tsunamis whose shape is influenced by the varying depth of the ocean \cite{degueldre_tsunami2016}.  

In Sec.\ \ref{sec:Description_sample}, we describe the chosen sample geometry, the choice of random disorder configurations, and the calculation of transport data. The application of an ANN that determines the amplitude and the correlation length of the disorder potential is presented in Sec.\ \ref{sec:2parameters}. In Sec.\ \ref{sec:fullpotential} we present two methods that allow to determine the full potential landscape from transport data, before we present our conclusions in Sec.\ \ref{sec:conclusions}. Appendix \ref{app:neural_net} gives details about the architectures of the ANNs which we use in this work.

\section{Description and generation of the samples}
\label{sec:Description_sample}

\subsection{Sample characteristics}
\label{sec:Sample_characteristics}

\begin{figure}
\centering
\includegraphics[width=0.9\linewidth]{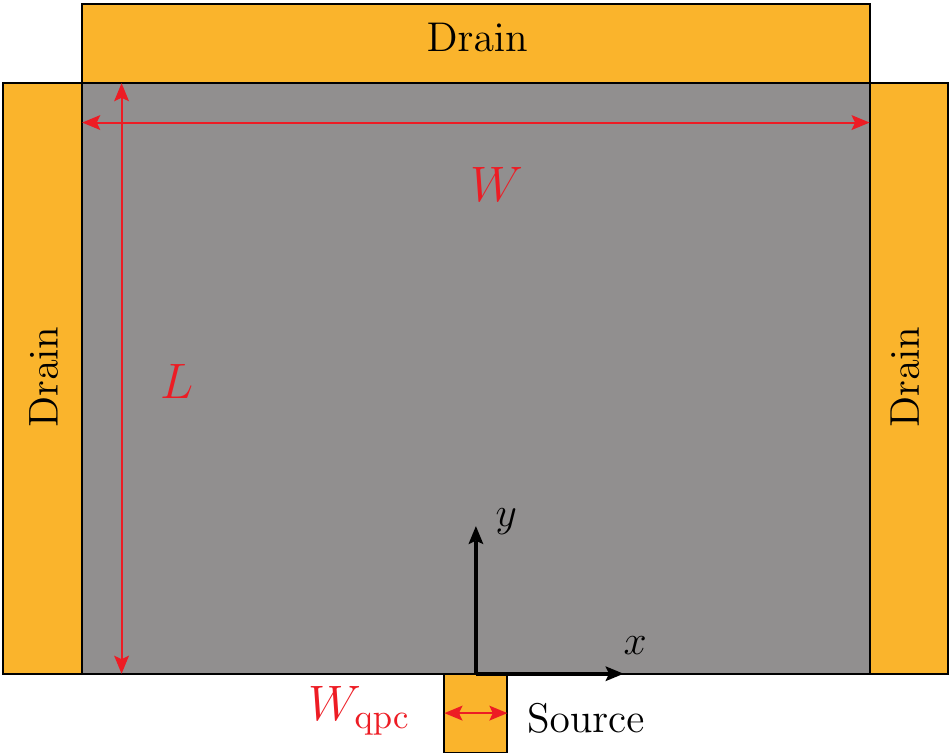}
\caption{Sketch of the system geometry used for the simulations. The quantum point contact is simulated by the small source electrode.}
\label{fig:sample}
\end{figure}
In this study, we focus on a two-dimensional electron gas (2DEG) arising in a semiconductor heterostructure. We assume a modulation-doping structure, where the impurities are located on a plane parallel to the 2DEG at a distance $s$, with a density $N_{\mathrm{d}}$. The heterostructure is connected to source and drain electrodes, and a voltage difference leads to an electron flow. In addition to the disorder potential of ionized dopants, we assume that the electrons are also subject to an electrostatic potential which creates a quantum point contact (QPC), see Fig.~\ref{fig:sample}.

In order to perform the transport simulations, we use the fully coherent tight-binding model quantum transport approach implemented through the Kwant package \cite{KWANT_2014}. The source is placed at the constriction of the QPC, and we focus only on the electrons injected through the QPC into a rectangular region (of dimension $L\times W = \unit[1.28] {\mu m} \times \unit[0.96] {\mu m}$) between the QPC with a width $W_{\mathrm{qpc}}= \unit[100] {nm}$ and the drain as represented on Fig.\ \ref{fig:sample}. In our simulations, we fix the lattice parameter to $a = \unit[5]{nm}$. We take values corresponding to GaAs/AlGaAs 2DEGs for the dielectric constant $\epsilon = 12.9$ and the effective mass $m^{*} \simeq \unit[5.74 \times 10^{-32}]{kg}$. The fixed electron Fermi energy $E = \unit[5.6]{meV}$ corresponds to a Fermi wave-length of $\lambda = \unit[65.5]{nm}$, much larger than $a$, and places us on the third conductance plateau of the QPC.

The only difference between samples is in the disorder potential. Experimentally, only two quantities can be chosen: the density of dopants $N_{\mathrm{d}}$ and the distance $s$ between the 2DEG and the dopant layer. These quantities define the global properties of the disorder potential like its strength and its correlation length. The precise realization of the dopant positions yields the individual disorder configuration. The disorder potential due to the Coulomb potential of the ionized dopants can be implemented following the approach of Ref.\ \cite{ihn2010semiconductor}. Using this method, the fluctuating potential is decomposed into a Fourier series, where the coefficents $C(\vec{q}_j)$ are complex numbers which follow a Gaussian distribution with a standard deviation $\sigma^2 = N_{\mathrm{d}}LW/2$. For a discrete lattice, one can write the disorder potential as \cite{ihn2010semiconductor}
\begin{equation}
\label{eq:randompotential}
    V(\vec{r}) = -\frac{\Delta q_x \Delta q_y}{\pi} \sum_{j \ne 0} \frac{e^{-q_j s}}{q_j + q_{\mathrm{TF}}} C(\vec{q}_j) e^{-i \vec{q}_j \cdot \vec{r}}, 
\end{equation}
where $\vec{r}$ is a two-dimensional vector in the plane of the 2DEG which denotes the position, 
the  discretization  of  the $\vec{q}$-space  gives  the  step  widths $\Delta q_{x} = 2\pi/L$ and $\Delta q_{y}=2\pi/W$, and $\vec{q}_j$ is the vector in the reciprocal space. The method described above takes also into account the Thomas-Fermi screening through the term $q_{\mathrm{TF}}$ which corresponds to the inverse of the effective Bohr radius. The  maximum $q$-values that we would have to consider are in principle given by $q_{\mathrm{max}} = 2\pi / a$. However, the decaying exponential allows us to neglect contributions from large $q$-values in the sum, and to evaluate $V(\vec{r})$ with reasonable computational effort. The chosen cutoff is $q_{\mathrm{cutoff}}= 3.5/s$.

\subsection{Transport simulations}
\label{sec:Simulation}

\begin{figure*}
\centering
\includegraphics[width=0.8\linewidth]{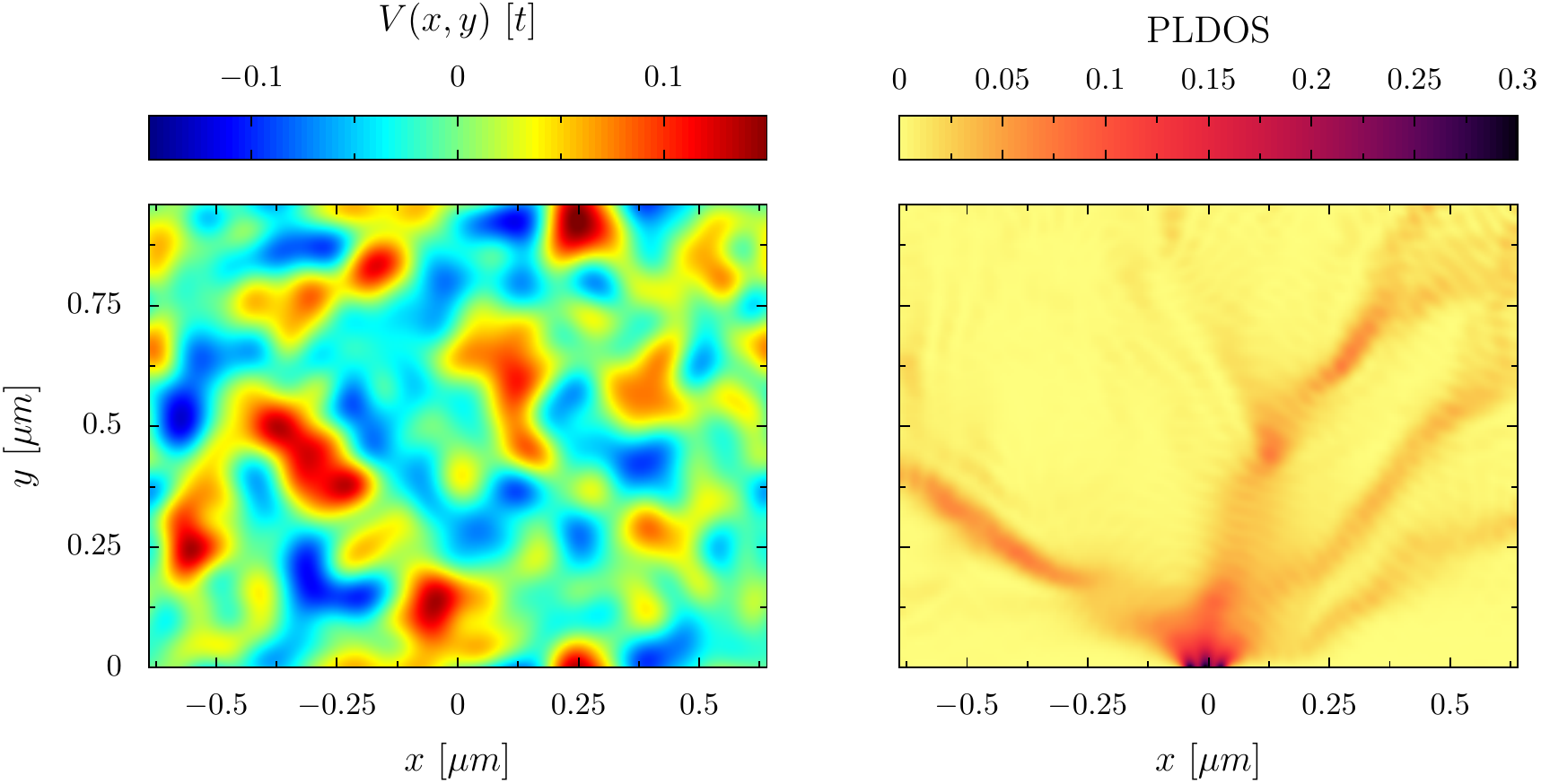}
\caption{Example of a random Coulomb potential in units of the hopping amplitude $t$ (left panel) and the corresponding calculated PLDOS (right panel), for a density $N_{\mathrm{d}} = \unit[10^{12}]{cm^{-2}}$ and a distance $s = \unit[50]{nm}$.}
\label{fig:pldos_pot}
\end{figure*}
To train an ANN, we need a dataset containing transport properties for various disorder configurations. Our neural networks are trained with supervised learning, which means that the training data (the transport properties) are associated to their corresponding target (the set of parameters we want to extract to characterize the disorder potential of the sample). The large amount of data required to create the dataset would not allow the use of experimental data. Moreover, the precise disorder configuration is usually not experimentally accessible, with the notable exception of light propagation in soap films \cite{patsyk20}. Nevertheless, we still want to characterize our sample from data that are in principle experimentally measurable. Then, a good candidate to fulfill this condition would be the SGM signal which is also easy to compute, at least at zero temperature. Since we also have to take into account the time required to produce the dataset composed of tens of thousands of samples, here we work with a less time-consuming transport property, which is the partial local density of states (PLDOS) \cite{buettiker1996,gramespacher1999}. 

The PLDOS corresponds to the contribution of the scattering states impinging from a lead to the local density of states, such that its structure describes the flow of electrons entering the sample through that lead. We work with the PLDOS from the QPC which reads
\begin{equation}
    \rho_{\mathrm{qpc},E}(\vec{r}) = 2 \pi \sum^{N_{\mathrm{qpc}}}_{n=1} |\psi_{\mathrm{qpc},E,n}(\vec{r})|^2,
\end{equation}
where $\psi_{\mathrm{qpc},E,n}(\vec{r})$ is the scattering wave-function of an electron injected from the channel $n$ of the QPC with an energy $E$, and the sum runs over the $N_{\mathrm{qpc}}$ open channels. 
The PLDOS, shown in the right panel of Fig.\ \ref{fig:pldos_pot}, is not experimentally accessible. It has however been shown that the SGM signal can be linked to the PLDOS at the Fermi energy, in particular at low temperatures and in the case of perfect transmission (e.g., if the energy corresponds to a conductance plateau) and for very weak disorder \cite{Ly2017}. Moreover, for a smooth and weak potential, the PLDOS is closely related to the density of classical trajectories starting from the QPC \cite{fratus2019,Keith_2021}. The branching pattern of the PLDOS depends sensitively on the disorder configuration. While we do not know how to solve the inverse problem of the transport calculation from the disorder potential, we show in the next sections that deep learning algorithms can extract the desired information about the disorder from the PLDOS.

\section{Extraction of two global parameters of the disorder}
\label{sec:2parameters}

Before using ANNs to recover the full potential landscape, we first focus on a simpler task: the characterization of the disorder with only two global parameters. The aim of this first step is to go beyond the simple application of neural networks and allows us to get a better understanding of the working procedure of the neural network. 

We created a dataset of 72.000 samples. For each sample, we chose a random distance $s$ between the dopant layer and the 2DEG between $40$ and $\unit[70]{nm}$, and a random dopant density $N_{\mathrm{d}}$ between $0.6 \times 10^{12}$ and $\unit[1.5 \times 10^{12}]{cm}^{-2}$. For those values, realistic for high-mobility samples, the amplitude of the disorder potential remains well below the Fermi energy. Then all the Fourier coefficients in Eq.\ \eqref{eq:randompotential} are randomly generated following a normal distribution with a standard deviation $\sigma^2 = N_{\mathrm{d}}LW/2$. Thus all the samples have a different disorder configuration. The left panel of Fig.\ \ref{fig:pldos_pot} shows an example. The dataset is composed of the PLDOS of the sample (the right panel of Fig.\ \ref{fig:pldos_pot}) and the associated targets which correspond to the set of global disorder parameters $(s,N_{\mathrm{d}})$.

\begin{figure*}
\centering
\includegraphics[width=0.75\linewidth]{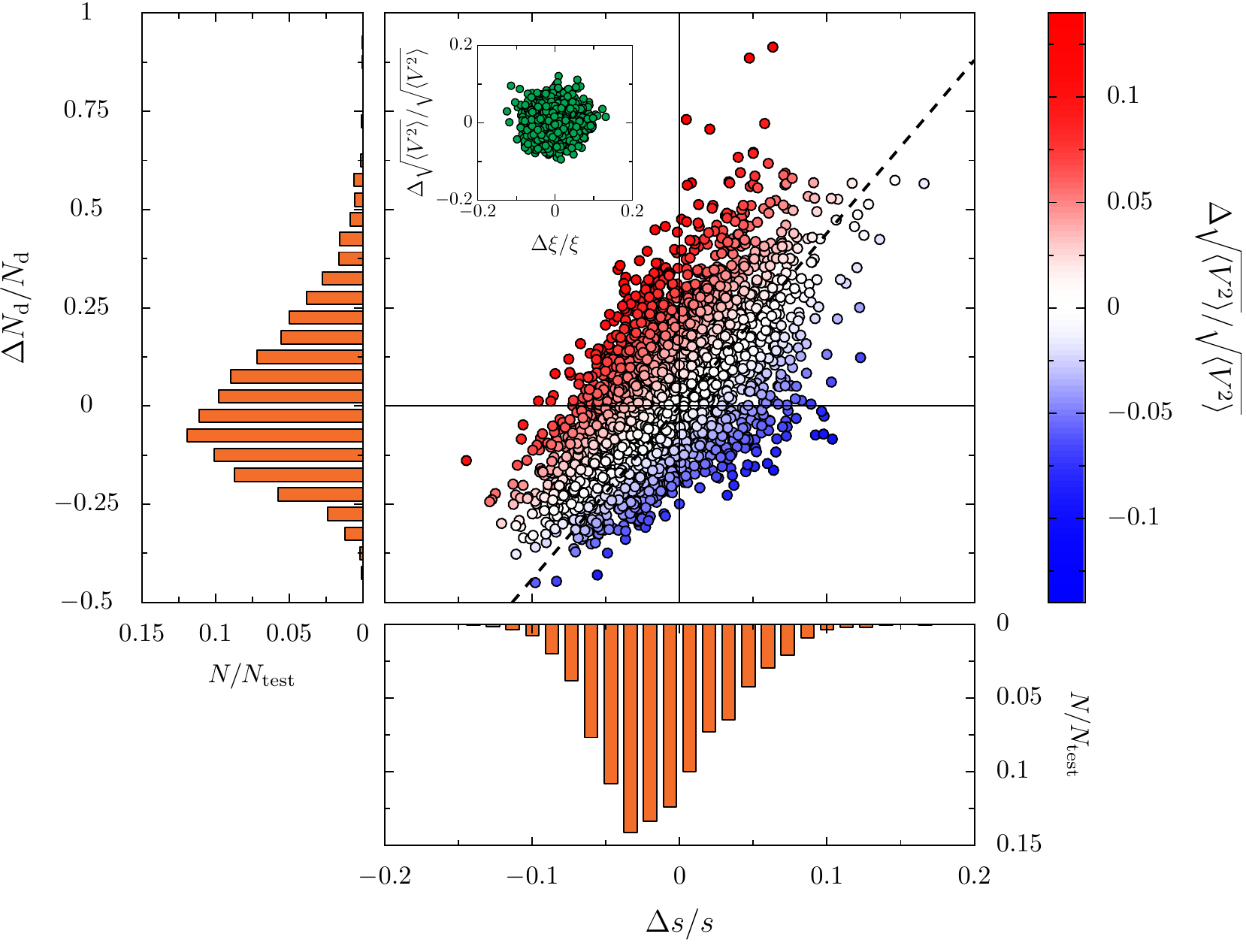}
\caption{Distribution of the relative errors $\Delta N_\mathrm{d}/N_\mathrm{d}$ of the dopant density and the distance $\Delta s/s$ for all elements of the test set scaled by the number of samples in the test set $N_{\mathrm{test}}$ (orange histograms). The central figure depicts the individual values of both errors and thereby puts in evidence the correlation between the two parameters. The color code indicates the relative error on the root mean square. The black dashed line represents the linearized theoretical line in the $\Delta N_{\mathrm{d}}/N_{\mathrm{d}}$ \textit{versus} $\Delta s/s$ plane for which the predicted root mean square corresponds to the expected root mean square. The inset represents the distribution of the relative errors of the potential root mean square $\Delta\sqrt{\langle V^2 \rangle}/\sqrt{\langle V^2 \rangle}$ and the correlation length $\Delta \xi/\xi$.}
\label{fig:Correlation_OP}
\end{figure*}
As the input of the neural network is an array (which corresponds to the PLDOS at each point of the two-dimensional space) similar to pixels of an image, the appropriate neural network architecture is a convolutional neural network (CNN) \cite{LeCun2015} \footnote{In this work we use an image which corresponds to a 2D array, but CNN can also handle 3D arrays.}. CNNs have the advantage to be composed of much less training parameters than densely connected neural networks. The output of the CNN is composed of two real numbers. In a first step, we ignore that $s$ and $N_{\mathrm{d}}$ both affect the strength of the potential landscape, and train the neural network such that the output is expected to correspond to those two parameters. The exact architecture of the CNN used in our study is described in App.\ \ref{app:DL2P}. 

Before the training process, the neural network is initialized, which means that the values of the network parameters are chosen randomly following a normal distribution. These parameters are modified during the training process, ideally with a convergence to values that minimize the deviation of the prediction (which depends on the ensemble of parameters of the neural network) from the target values. The deviation is quantitatively evaluated through the loss function, and the convergence process corresponds to finding a global minimum of this loss function. It is then clear that due to the complexity of the network, the convergence depends critically on the initial values of the parameters. 
However, when training the CNN to recover the set of targets $(s,N_{\mathrm{d}})$, we notice a clear regularity in the accuracy of the models to predict these quantities on the test set \footnote{The test set is an ensemble of samples similar to the ones which compose the training set. However, they are not used in the training process.}. We find a standard deviation of the prediction error for the distance $\sigma_s = \unit[2.3]{nm}$  and $\sigma_{N_{\mathrm{d}}} = \unit[1.8 \times 10^{11}]{cm}^{-2}$ for the density (typical value of $s$: $\unit[50]{nm}$, typical value of $N_{\mathrm{d}}$: $\unit[10^{12}]{cm}^{-2}$). The difference of precision of these two parameters can be explained by their different impacts on the disorder configuration. Such impacts can be quantified by using the potential root mean square $\sqrt{\langle V^2 \rangle}$ and the correlation length $\xi$ of the disorder. The latter quantities are defined from the auto-correlation function of the potential
\begin{equation}\label{eq:potential_autocorrelation}
    \mathcal{C}(\Vec{R}) = \langle V(\Vec{r}) V(\Vec{r} + \Vec{R}) \rangle.
\end{equation}
For the potential given in Eq.\ \eqref{eq:randompotential}, the correlation function reads
\begin{equation}
    \mathcal{C}(\Vec{R}) = 2 \frac{\sigma^2}{\pi^2} \int_{0}^{q_{\mathrm{cutoff}}} \mathrm{d}q\ q \frac{e^{-2q s}}{(q + q_{\mathrm{TF}})^2}\ J_0(qR),
\end{equation}
where $J_0(z)$ is the zero order Bessel function of the first kind. The correlation length $\xi$ is defined such that $\mathcal{C}(\xi)/\mathcal{C}(0) = 1/2$. The potential root mean square is given by $\sqrt{\langle V^2 \rangle} = \sqrt{\mathcal{C}(0)}$ and thus reads
\begin{equation}
    \sqrt{\langle V^2 \rangle} = 2 \frac{\sigma^2}{\pi^2} \int_{0}^{q_{\mathrm{cutoff}}} \mathrm{d}q\ q \frac{e^{-2q s}}{(q + q_{\mathrm{TF}})^2}. \label{eq:PRMS}
\end{equation}

Figure \ref{fig:Correlation_OP} shows the relative errors in $s$ and $N_{\mathrm{d}}$ for the elements of the test set. One can see that in most cases the distance and the density are either both overestimated or both underestimated. This result is not surprising since an overestimation of the density leads to an overestimation of the potential root mean square, which is compensated by overestimating the distance (an overestimation of the distance leads to an underestimation of the potential root mean square). The hypothesis that even if the target parameters are the distance and the density, the CNN internally determines the potential root mean square and the distance $s$, and then tries to determine the density $N_{\mathrm{d}}$ with these two quantities, explains the correlation between the two relative errors observed in Fig.\ \ref{fig:Correlation_OP}. Along the dashed black line, the lowest order corrections due to the relative errors $\Delta N_{\mathrm{d}}/N_{\mathrm{d}}$ and $\Delta s/s$, evaluated at the mean value of $s$, compensate each other in the formula for the potential root mean square \eqref{eq:PRMS}. One can see that the correlation of the errors of prediction of the model follows that line on which the real potential root mean square is close to the value stemming from the predicted parameters. One can also notice that the samples with an overestimated (underestimated) density are the samples with a density close to the lower (upper) bound which corresponds to $N_{\mathrm{d}} = \unit[0.6 \times 10^{12}]{cm}^{-2}$ ($N_{\mathrm{d}} = \unit[1.5 \times 10^{12}]{cm}^{-2}$). This issue could be due to the finite range of parameters used in the training set. As we train the neural network with samples with a uniform distribution of density in a given range, the neural network has a higher probability to predict a density in the expected range even for a sample with a density close to the bound. The same phenomenon occurs for the distances but it is less pronounced.

While the global parameters $s$ and $N_\mathrm{d}$ both affect the potential strength, it is more natural to characterize the potential landscape by parameters that independently describe its amplitude and its roughness. Such independent parameters are the root mean square $\sqrt{C(0)}$ and the correlation length $\xi$ of the potential. 
Interestingly, when training a CNN to directly recover these two independent quantities, a better characterization of the potential is possible (see the inset of Fig.\ \ref{fig:Correlation_OP}). The standard deviation of the prediction error for the potential root mean square is $\sigma_{\mathrm{PRMS}} = 7.3 \times 10^{-3} t$ and the standard deviation of the prediction error for the correlation length is $\sigma_\xi = \unit[0.36]{nm}$ (typical value of the potential root mean square: $0.25 t$, typical value of $\xi$: $\unit[10] {nm}$). Then a precise characterization of the disorder with two parameters appears to be possible. We also notice that the distance $s$ and the correlation length $\xi$ can be considered as proportional for $\unit[40]{nm} \leq s \leq \unit[70]{nm}$. Then the accuracy on the distance can be slightly improved by performing the regression on the correlation length and then deducing the distance $s$. 

\section{Determination of the full potential landscape}
\label{sec:fullpotential}

While the previous characterization of the disorder by two global parameters is very useful, it is also possible to extract more precise information using ANNs, and in particular to determine the full potential landscape of Eq.\ \eqref{eq:randompotential}, created by the dopants, from the knowledge of the PLDOS. This can be achieved by two different methods. The first one consists in choosing the set of Fourier coefficients as target of the ANN. The second one uses a neural network of the type called convolutional encoder-decoder which has the particularity of taking an image as input and to give another image as output. In our case the targeted image corresponds directly to the 2D array of the values of the potential landscape in real space.

\subsection{Potential reconstruction through its Fourier coefficients}
\label{sec:fullfourier}

\begin{figure}
\centering
\includegraphics[width=\linewidth]{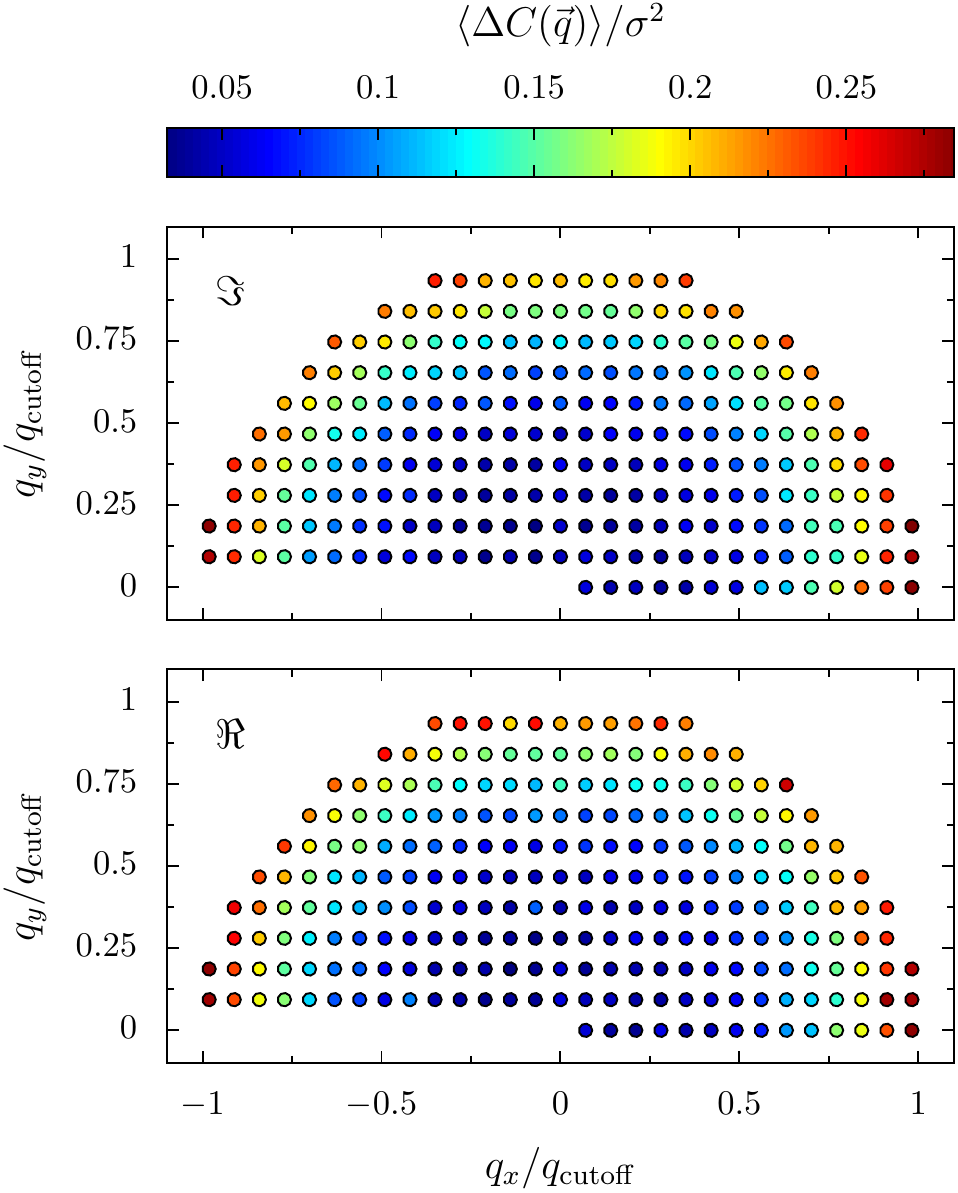}
\caption{Mean square of the error $\Delta C(\vec{q})$ in the predicted Fourier coefficients (scaled by $\sigma^2$), evaluated using 3000 test samples. The points correspond to the Fourier coefficient and are localized at their associated $\vec{q}$-vector. The top (bottom) panel represents the imaginary (real) part of the Fourier coefficients. Only half of the Fourier coefficients are depicted in this figure, the others are given by their complex conjugate values.}
\label{fig:error_FC}
\end{figure}
\begin{figure*}
\includegraphics[width=0.65\linewidth]{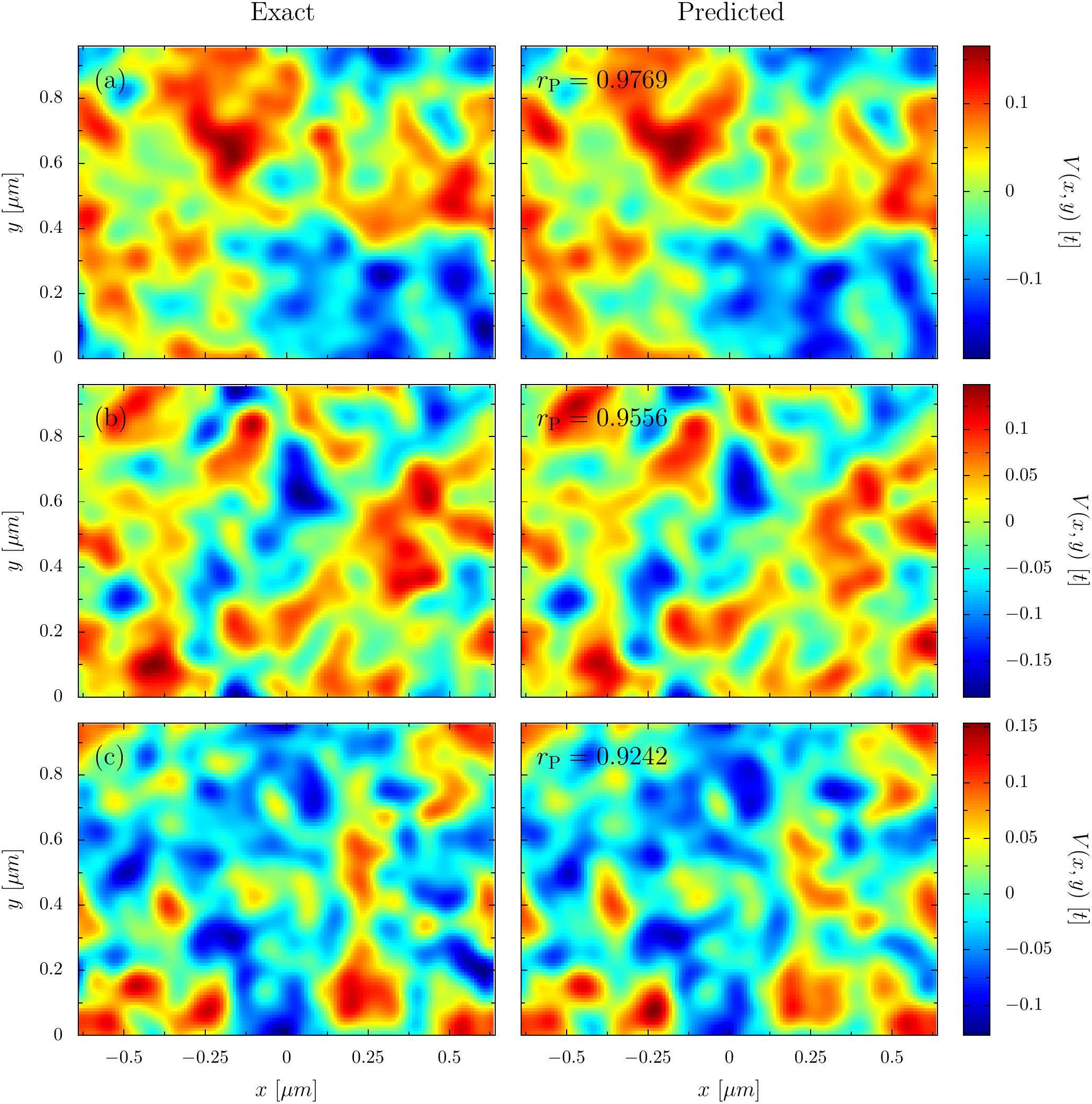}
\caption{Example of potential landscapes reconstructed through the regression on the Fourier coefficents. The left (right) panel corresponds to the expected (predicted) potential. The correlation coefficients $r_\mathrm{P}$ of the predicted potentials with the exact ones are evaluated using Eq.\ \ref{eq:pearson} and correspond to the value range indicated in the left panel of Fig.\ \eqref{fig:hist_error} by the associated letter.}
\label{fig:example_reconstruction_FC}
\end{figure*} 
Let us first describe the method which aims at determining the set of Fourier coefficients. This method assumes that the potential is given in the form of Eq.\ \eqref{eq:randompotential} in terms of the Fourier coefficients, the distance $s$ between the dopant layer and the 2DEG, and the value of the Thomas-Fermi constant. Once the Fourier coefficients are determined by the ANN, we use Eq.\ \eqref{eq:randompotential} to find the real-space potential of the sample.

In order to train and test this method, we created a dataset containing 88.000 samples. Those samples have the characteristics described in Sec.\ \ref{sec:Sample_characteristics}, with a fixed distance $s = \unit[50]{nm}$ and a fixed density at $N_{\mathrm{d}} = \unit[10^{12}]{cm}^{-2}$. The precise architecture of the neural network is presented in App.\ \ref{app:DLFC}. The architecture of the neural network depends on the number of Fourier coefficients to extract $\mathcal{N}$, which in turn depends on the distance $s$ and the size of the system. Thus, the smaller $s$, or the larger the sample, the more coefficients will have to be determined. Obviously, the size of the output layer is determined by the number $\mathcal{N}$ of Fourier coefficients we want to find. The number of neurons on the second last layer should also be adapted as well as the learning rate, the size of the batch and the number of filters in the convolution layer. To maintain a good reliability, the number of neurons has to increase with the size because in addition to the increasing number of Fourier coefficients, they will also correspond to closer spatial frequency values $\vec{q}$, which complicates the problem.
\begin{figure*}
\centering
\includegraphics[width=0.8\linewidth]{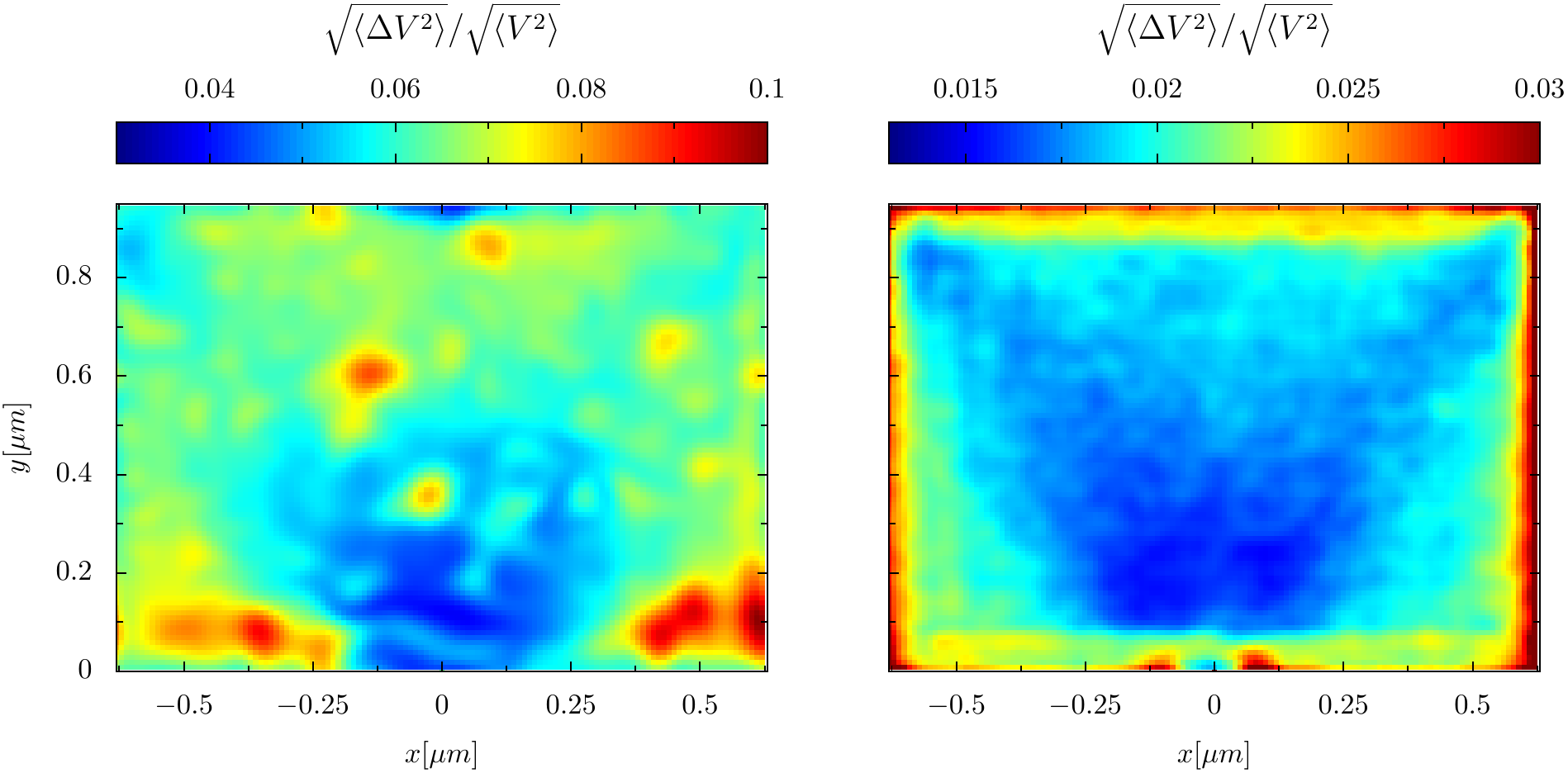}
\caption{Root of the average squared local error in the real space potential (scaled by the root mean square of the potential used for the generation of samples), calculated from over 3000 test samples. The left and right panel corresponds to the methods of the Fourier coefficients and the encoder-decoder, respectively.}
\label{fig:error_realspce}
\end{figure*}
\begin{figure*}
\centering
\includegraphics[width=0.65\linewidth]{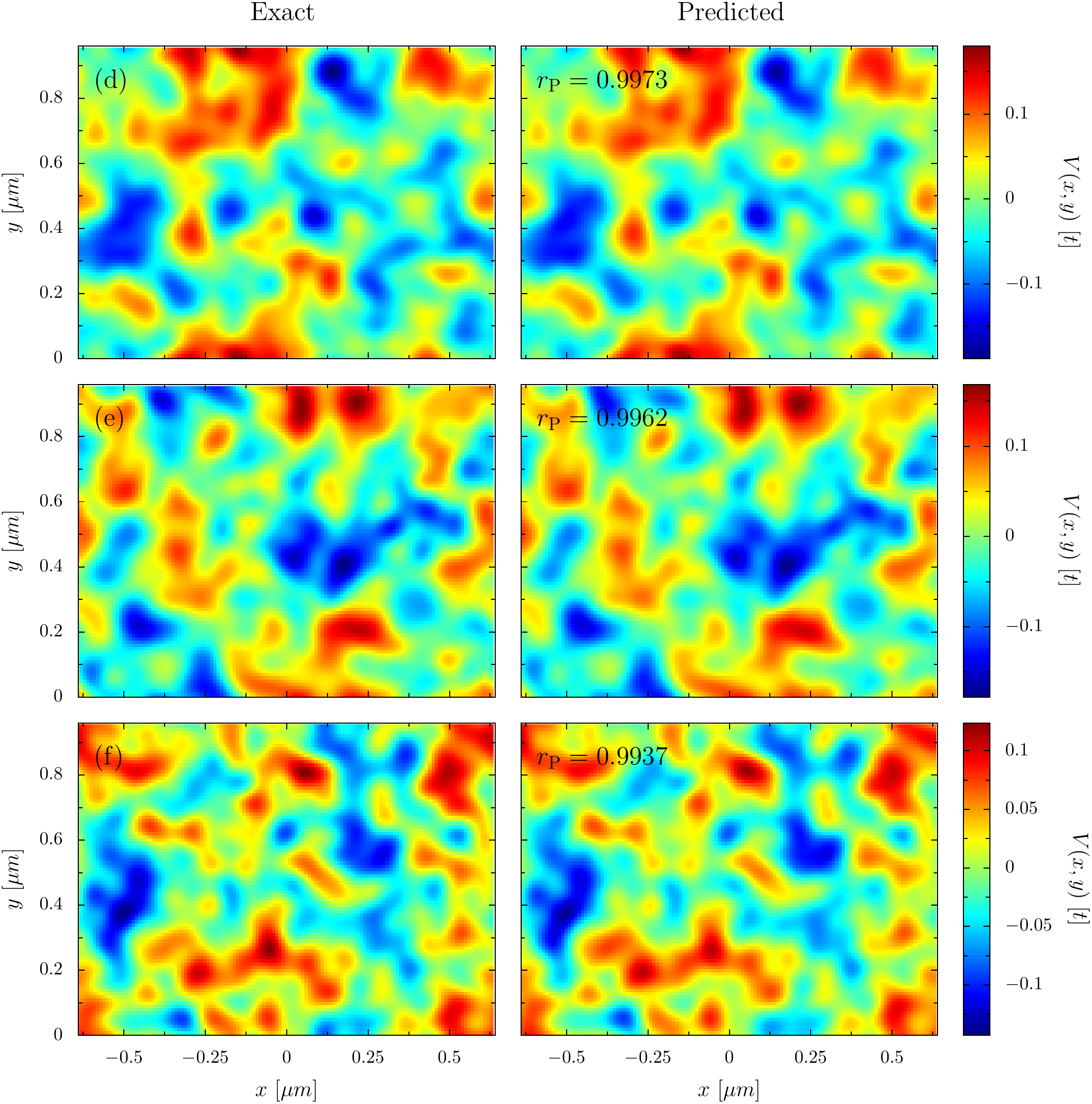}
\caption{Examples of potential landscapes reconstructed with the direct method based on a convolutional encoder-decoder network. The left (right) panel corresponds to the expected (predicted) potential. The correlation $r_\mathrm{P}$ of the predicted potentials with the exact ones is quantified through Eq.\ \eqref{eq:pearson} and corresponds to the value ranges indicated in the right panel of Fig.\ \ref{fig:hist_error} by the associated letter.}
\label{fig:example_reconstruction_ED}
\end{figure*} 

Before evaluating the quality of the potential reconstruction, we first focus on the accuracy of the model prediction for the Fourier coefficients. As depicted in Fig.\ \ref{fig:error_FC}, we observe that the model is rather reliable for small absolute values of $q$, and that it has more difficulties to predict coefficients corresponding to large $q$, with a few exceptions that depend on the model. This result is plausible, knowing that the Fourier coefficients are weighted by a decreasing exponential with the norm of $q$. Thus, the Fourier coefficients corresponding to a large $q$ have a less important impact on the potential landscape than Fourier coefficients associated to a small $q$. 

\begin{figure*}
\centering
\includegraphics[width=0.8\linewidth]{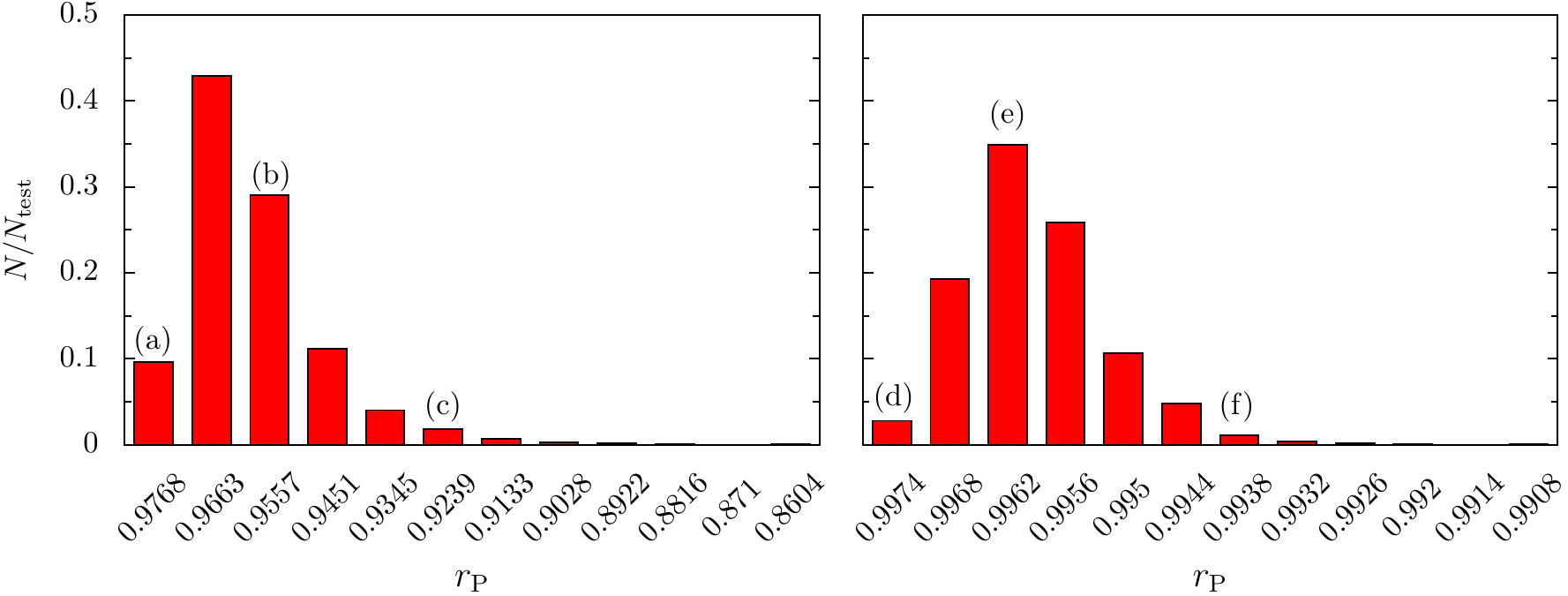}
\caption{Distribution of the Pearson correlation coefficient $r_\mathrm{P}$ for the samples of the test set, scaled by the number of coefficients in that set. The left (right) panel corresponds to the indirect Fourier coefficient (direct method based on a convolutional encoder-decoder network). The letters are used to label examples shown in Figs.\ \ref{fig:example_reconstruction_FC} and \ref{fig:example_reconstruction_ED} which are situated in the corresponding interval of error.}
\label{fig:hist_error}
\end{figure*} 
Examples of reconstructed potential landscapes in real space are shown in Fig.\ \ref{fig:example_reconstruction_FC}. To analyze the reconstruction of the potentials, the error distribution in real space, averaged over the full test set, is shown in the left panel of Fig.\ \ref{fig:error_realspce}. In regions where the PLDOS assumes large values, as it is typically the case close to the QPC, the potential is reproduced with the highest accuracy. Though the Fourier coefficients have the same contribution in all regions of real  space, their values seem to be determined by the neural network from the regions where the structure of the PLDOS is most significant.

\subsection{Direct determination of the real-space potential}
\label{sec:fulldirect}

The second method is more straightforward since the output of the neural network is directly the real-space potential landscape. This implies that we do not need any supplementary information than the PLDOS (input of the neural network) to reconstruct the full potential landscape. The exact architecture of the convolutional encoder-decoder neural network used is described in App.\ \ref{app:ED}. As in the previous case, the accuracy of the model prediction depends on the size of the input and output image but the dependency is less pronounced. The convolutional encoder-decoder network is trained using the same dataset as in the case of the previous method.

Three examples of such a direct disorder landscape reconstruction by the convolutional encoder-decoder network are shown in Fig.\ \ref{fig:example_reconstruction_ED}.  
As shown in the right panel of Fig.\ \ref{fig:error_realspce}, the error of the potential prediction averaged over the test set is smaller in the center of the sample. This could be due to the fact that the partial density of states is, in average, more important in this region. We however notice a relatively large error in front of the QPC when using the convolutional encoder-decoder network, despite a large PLDOS in that region. We assume that such a behavior is due to the strong density of electrons in front of the QPC, independently of the disorder, and the fact that important disorder-dependent features like branches in the PLDOS only develop at a certain distance from the QPC \cite{fratus2019,patsyk20}. Enhanced errors at the borders of the sample could be due to the fact that a precise estimation of a weak scattering potential necessitates the input of the electron flow behind the region of interest.

\subsection{Evaluation and comparison of the two methods}

The examples of reconstructed potential landscapes presented in Figs.\ \ref{fig:example_reconstruction_FC} and \ref{fig:example_reconstruction_ED} show qualitatively that a reliable determination of the potential is achieved. From the different scales in Fig.\ \ref{fig:error_realspce}, we already observe that using the convolutional encoder-decoder based direct method is in average much more accurate than the method which reconstructs the potential through a determination of the Fourier coefficients. In order to compare the predicted potential to the expected one in a more quantitative way, we use the Pearson correlation coefficient
\begin{equation}\label{eq:pearson}
r_\mathrm{P} = \frac{\sum^{N}_{i=1} (y_{\mathrm{pred},i} - \bar{y}_{\mathrm{pred}})(y_{\mathrm{exp},i} - \bar{y}_{\mathrm{exp}})}{\sqrt{\sum^{N}_{i=1} (y_{\mathrm{pred},i} - \bar{y}_{\mathrm{pred}})^2}\sqrt{\sum^{N}_{i=1} (y_{\mathrm{exp},i} - \bar{y}_{\mathrm{exp}})^2}},
\end{equation}
where the summations run over the $N$ pixels of the image. 
$\bar{y}_{\mathrm{pred}}$ and $\bar{y}_{\mathrm{exp}}$ are the average values of the predicted potential and the expected potential, respectively. The value of $r_\mathrm{P}$ approaches $r_\mathrm{P}=1$ in the case of perfectly correlated images. 

\begin{figure*}
\centering
\includegraphics[width=0.7\linewidth]{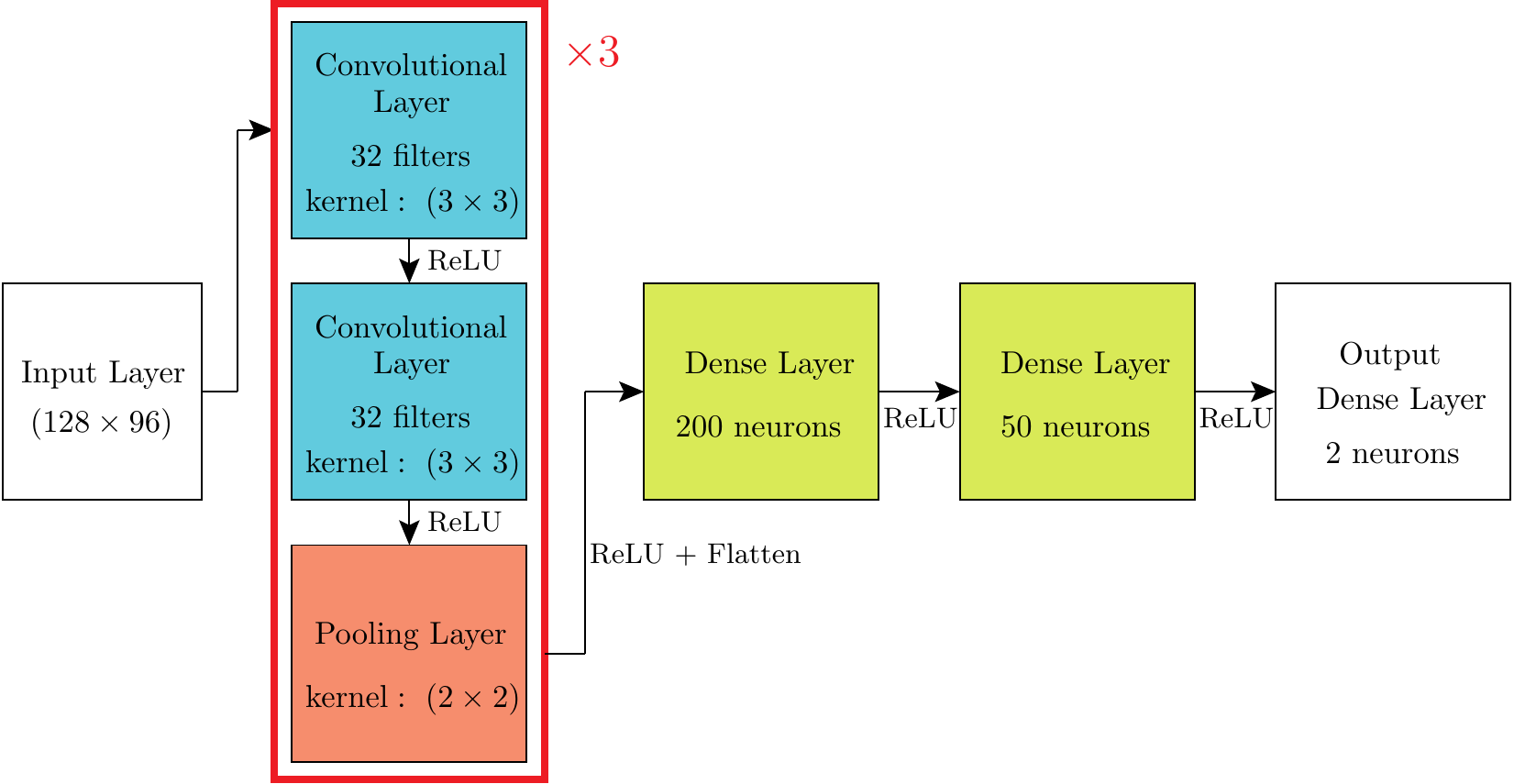}
\caption{Architecture of the convolutional neural network used for the regression on two parameters. The three layers in the red rectangle are repeated three times. The activation function used between all the layers is the ReLU function \eqref{eq:relu}.}
\label{fig:CNN_2P}
\end{figure*}
The distribution of $r_\mathrm{P}$ for the reconstruction of the potential of the samples contained in the test set is depicted in Fig.\ \ref{fig:hist_error}. Though the correlation is high in all cases, the direct results of the convolutional encoder-decoder network (right panel) are considerably more accurate (with $r_\mathrm{P} > 0.99$ and thus almost perfect correlation in all cases) than the potential reconstructed from the Fourier coefficients (left panel; typically $r_\mathrm{P}\approx 0.95$, and rare cases with $r_\mathrm{P}<0.9$). The examples from the test set shown in Figs.\ \ref{fig:example_reconstruction_FC} and \ref{fig:example_reconstruction_ED} are labeled by letters which correspond to the ones in Fig.\ \ref{fig:hist_error}, where they are associated to an error interval. For both methods, the shown examples are for one of the best reconstructions (a, d), an average precision (b, e), and one of the worst (c, f).

 While the direct convolutional encoder-decoder method is faster (see App.\ \ref{app:ED}) and more accurate, it is not restricted to provide a prediction of the potential which is supposed to correspond to a Coulomb potential arising from ionized dopants placed at a certain distance from the 2DEG \footnote{This kind of procedure using neural networks and physical equations ressembles to a physics-informed neural network which combines neural networks and non-linear differential equations to give a physically correct result \cite{Karniadakis2021}}. Moreover, the output of the convolutional encoder-decoder network gives the potential landscape with a resolution that corresponds to the resolution of the image of the targeted potential used during the training process. However, an important advantage of the convolutional encoder-decoder network is the direct prediction of the potential from the PLDOS, without any other information required. It could therefore be used to determine disorder landscapes in various systems like graphene or TMDs, where the origin of the disorder potential is not always known. Finally, we note that both methods give better results in the center of the sample.

\begin{figure*}
\centering
\includegraphics[width=0.7\linewidth]{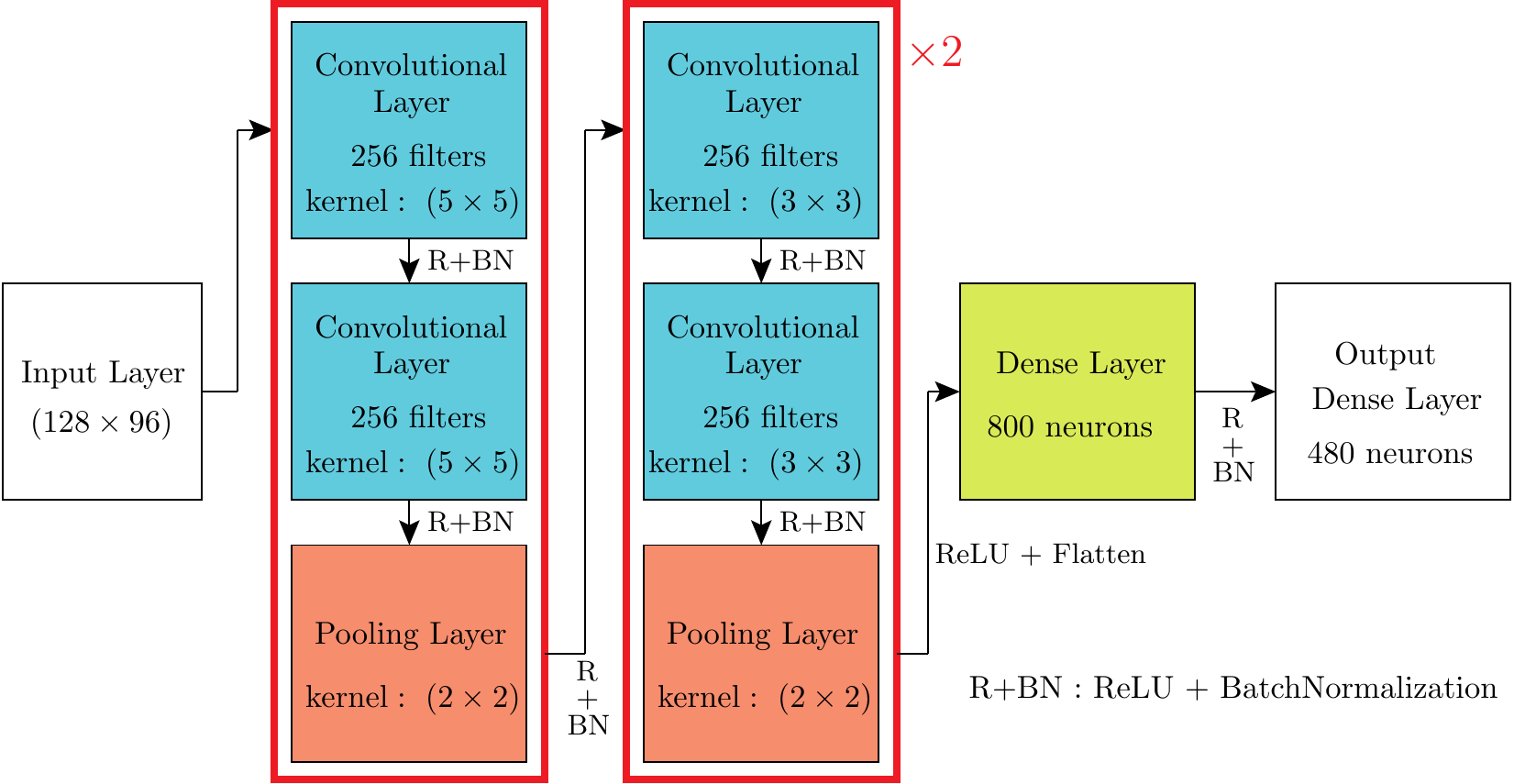}
\caption{Architecture of the convolutional neural network used for the regression on the Fourier coefficients. The three layers in the second red rectangle are repeated two times. Between all the layers we used the ReLU function and a batchnormalization (BN) layer.}
\label{fig:CNN_FC}
\end{figure*}
\section{Conclusions}
\label{sec:conclusions}

We have developed a machine-learning approach to extract properties of the potential landscape from electronic transport data. For the example of electrons injected through a QPC into a 2DEG, we have calculated the PLDOS (that can be related under certain conditions to measurable SGM data \cite{Ly2017}) for many different disorder realizations using Kwant \cite{KWANT_2014}, and used the set of data pairs to train CNNs to recognize different features of the disorder potential from a given PLDOS. We have shown that such an approach allows to extract the global properties of the potential like the typical amplitude and the correlation length. 

Our study with two different pairs of target parameters to characterize the disorder illustrates the important conclusion that the machine learning approach might be extremely useful, but it is not a substitute for the understanding that arises from the physical analysis of the problem. Without the latter, the two studied pairs of parameters would be equivalent until a detailed analysis of the correlations (like that of Fig.\ \ref{fig:Correlation_OP}) is done. But basic physical analysis of electron transport in disordered systems readily leads to the prediction that the second pair (correlation length and potential root mean square) has independent parameters. 

We have demonstrated that it is even possible to use an ANN to reconstruct the full potential landscape. To this end, we have trained a CNN to extract the Fourier coefficients of the potential landscape from PLDOS data, and demonstrated that the potential can be reconstructed with rather good accuracy. An alternative method using a convolutional encoder-decoder shaped ANN, trained with real-space potential data produces even more impressive results with an almost perfect reconstruction of the full potential landscape.

Our results demonstrate that it is possible to use machine learning to determine the potential landscape from electronic transport data. Such a method should be of interest in the field of nanoelectronics, where a better control of disorder-realization dependent properties and the resulting sample-to-sample fluctuations could be achieved once the disorder potential is accessible. Moreover, the presented method could be used to characterize the different disorder contributions in different two-dimensional conductors like 2DEGs and graphene. 

While the accuracy found in our study is very high, such a performance depends on the size of the set of training data. When using the SGM response as input, one has to compute the finite-temperature conductance (i.e., a weighted average of the transmission over energy) for many different tip positions. Even though the weak energy-dependence of electron flow \cite{braem2018,fratus2019} facilitates the energy integral, the creation of the training set will necessitate an enormous amount of numerical resources. However, the electron flow is affected by a magnetic field perpendicular to the 2DEG, and it is expected that the accuracy could be considerably improved (and therefore the number of samples needed for training reduced) when a disorder realization is related to the ensemble of transport data obtained at different magnetic field values \cite{percebois_in_progress}.

\acknowledgments
The authors are grateful to Thomas Allard, Jonathan Chardin, Keith Fratus, Mauricio G\'{o}mez Viloria, Riccardo Hertel, Rodolfo Jalabert, and Guillaume Weick for useful discussions.
This work of the Interdisciplinary Thematic Institute QMat, as part of the ITI 2021-2028 program of the University of Strasbourg, CNRS, and Inserm, was supported by IdEx Unistra (ANR 10 IDEX 0002), and by SFRI STRAT'US project (ANR 20 SFRI 0012) and EUR QMAT ANR-17-EURE-0024 under the framework of the French Investments for the Future Program. The authors would like to acknowledge the High Performance Computing Center of the University of Strasbourg for supporting this work by providing scientific support and access to computing resources. Part of the computing resources were funded by the Equipex Equip@Meso project (Programme Investissements d'Avenir) and the CPER Alsacalcul/Big Data.

\appendix

\section{Architectures of the Artificial Neural Networks}
\label{app:neural_net}

Deep neural networks \cite{deeplearning2016} are composed of a succession of neuron layers, typically with connections between neurons in adjacent layers. Each neuron and each connection contain trainable parameters. In order to find the best set of these parameters, we use the training data which are composed of the input of the neural network and the expected output. The training process consists in finding the values of those parameters which minimize the discrepancy between the prediction of the neural network and the targeted values \footnote{The discrepancy is evaluated quantitatively through a loss function. For the regression problem of the present section, the loss function we use is the squared error, averaged over the elements of the training set.}. For our large networks with many parameters, an exact determination of the global minimum of the loss function is impossible. Technically, the optimization of the set of parameters is then realized iteratively using a steepest-descent-like process. The so-called optimizer updates the parameters after seeing $N_{\mathrm{B}}$ elements of the training set where $N_{\mathrm{B}}$ is called the batch size. One of the most determinant parameters for the update of the parameters is the learning rate. Once a model is trained, its performance is evaluated on data from the test set that were not used in the training process. The shape and the parameters of the used networks have been determined empirically such that a good performance is obtained.

\begin{figure*}
\centering
\includegraphics[width=0.55\linewidth]{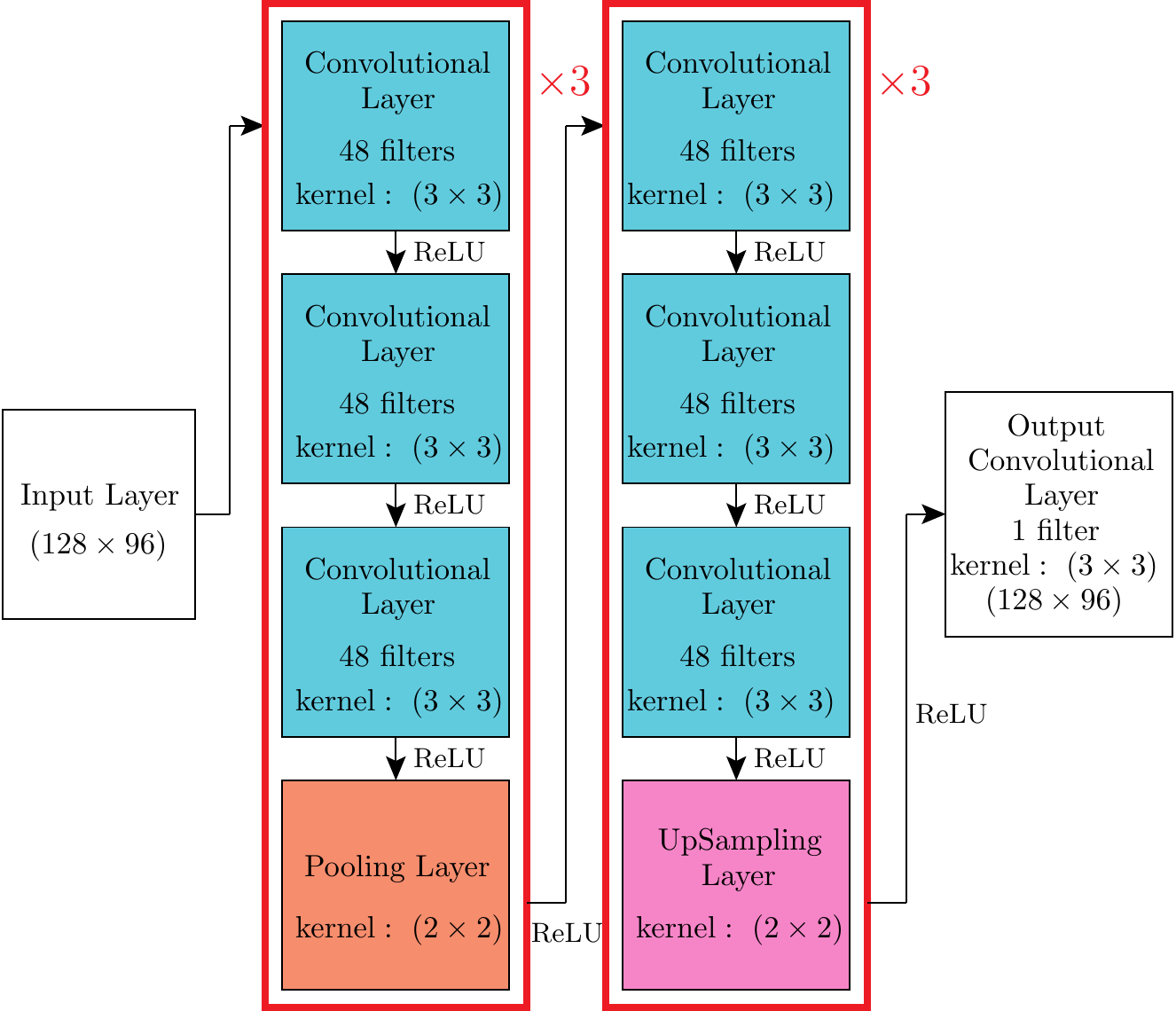}
\caption{Architecture of the convolutional encoder-decoder used for the direct reconstruction of the potential in real space. The encoder corresponds to the first block of four layers repeated three times and the decoder corresponds to the second block of four layers repeated three times. The ReLU function \eqref{eq:relu} is applied between all layers.}
\label{fig:CNN_ED}
\end{figure*} 
\subsection{Neural Network for two-parameter regression }
\label{app:DL2P}
In our case, the input of the neural network is always an image-like 2D array which corresponds to the position-dependent PLDOS values. An appropriate architecture for this kind of problem is a neural network that is divided in two parts. The first part is the convolutional analysis of the image which is composed of two kinds of layers: the convolutional layer \footnote{The convolutional layer applies filters on the previous images in order to detect particular features in the images. The main characteristics of a convolution layer is the size of the filters and the number of filters.} and the pooling layer \footnote{The aim of the pooling layer is to reduce the dimension (i.e., the height and the width) of the previous layer in order to keep only the relevant information. This reduction condenses the values of a small cluster of pixels of the previous layer in a single value. A pooling layer is characterized by the number of pixels which compose the cluster and the function applied on the cluster to reduce it to a single number. In our study we always use the MaxPooling which consists in taking the largest value of the cluster.}. The second part is composed of densely connected layers. Such a layer is characterized by the number of neurons which compose it. An activation function is applied on the output of each neuron in order to introduce non-linearity in the final output of the neural network, and the possibility to perform complex tasks. Between the convolutional part and the dense layers, we use a ``flatten layer'' which literally flattens the data of the convolutional layer into a one-dimensional vector.

The set of data $X$ are normalized in order to be more efficient during the training of the neural network \cite{aggarwal2018}. In our work, we use two kinds of normalization, the first one is the standardization
\begin{equation}\label{eq:standardization_normalization}
\hat{X} = \frac{X-\mu_{X}}{\sigma_{X}},
\end{equation}
where $\mu_{X}$ and $\sigma_{X}$ correspond to the average and the standard deviation of the data $X$, respectively. The second one is the Min-Max normalization
\begin{equation}\label{eq:minmax_normalization}
\hat{X} = \frac{X-\mathrm{min}_{X}}{\mathrm{max}_{X}-\mathrm{min}_{X}},
\end{equation}
where $\mathrm{min}_{X}$ and $\mathrm{max}_{X}$ is the minimum and maximum value of the data $X$.

Deep learning algorithms can be considerably accelerated when using the Graphical Processing Unit (GPU) of powerful graphics cards whose characteristics are well adapted to the numerical tasks due to their large number of cores which allows an efficient parallelization.

The architecture of the convolutional neural network we use in Sec.\ \ref{sec:2parameters} to perform the regression on two parameters from a 2D array representing the PLDOS is depicted in Fig.\ \ref{fig:CNN_2P}. We first repeated three times the block composed of two convolutional layers and one maxpooling layer for the convolutional analysis, and then we used two dense layers before the output layer. The chosen activation function is the Rectified Linear Unit (ReLU) function 
\begin{equation}\label{eq:relu}
x \rightarrow \left\{ 
\begin{array}{lcl}
0 & \mathrm{for} & x<0 \\ 
x & \mathrm{for} & x>0 
\end{array} \right. ,
\end{equation}
which is among the ones that lead to the best results. For the training, we used the well-known optimizer called Adam \cite{Adam_2014} with a learning rate of $10^{-3}$. The loss function used is the mean squared error. We set the batch size to $N_{\mathrm{B}}= 32$ and we performed the training over 5 epochs \footnote{The number of epochs corresponds to the number of times that the entire training set is used to adjust the parameters of the neural network.}. For this study, we used the standardization normalization \eqref{eq:standardization_normalization}. For all the other parameters we kept the ones which are set by default in the package Keras \cite{chollet2015keras}. The training process took a few minutes with the use of an Nvidia P100 GPU.

\subsection{Regression on Fourier coefficients}
\label{app:DLFC}

The CNN used in Sec.\ \ref{sec:fullfourier} to find the Fourier coefficients is shown in Fig.\ \ref{fig:CNN_FC}. It differs from the one described in App.\ \ref{app:DL2P} since the number of targeted parameters is much larger. The first difference is the size of the filter for the first convolutional block which is $(5 \times 5)$. Due to some issue in the training process, we used a very low learning rate of $10^{-5}$, and a higher batch size of 128. We performed the training over 10 epochs which was needed and sufficient to converge. The last difference is that we used only one dense layer of 800 neurons before the output layer whose size is given by twice the number of complex Fourier coefficients. As in the network described in  App.\ \ref{app:DL2P}, we use the standardization normalization \eqref{eq:standardization_normalization}.
The training took about 2 hours on an Nvidia P100 GPU.

\subsection{Convolutional encoder-decoder architecture}
\label{app:ED}

The convolutional encoder-decoder neural network shown in Fig.\ \ref{fig:CNN_ED} is used in Sec.\ \ref{sec:fulldirect} to directly find an image from another image. The aim of the encoder is to reduce the size of the image, keeping only the most relevant information by using convolutional and pooling layers (we repeated three times a block of three convolutional layers and one pooling layer). The decoder reconstructs the image from the information given at the end of the encoder. This reconstruction is performed by using convolutional and UpSampling \footnote{The UpSampling layer is the inverse of a pooling layer.} layers (we repeated three times a block of three convolutional layers and one UpSampling layer). At the bottleneck we have 48 feature maps with a size smaller than the original image. We used the Min-Max normalization \eqref{eq:minmax_normalization} and performed the training over 24 epochs with a batch size of $N_{\mathrm{B}} = 8$. As for the previous neural networks, we used the optimizer Adam with a learning rate of $10^{-3}$ and the mean squared error as the loss function. The training took about 1 hour on an Nvidia P100 GPU.

\bibliography{references} 

\begin{thebibliography}{38}%
\makeatletter
\providecommand \@ifxundefined [1]{%
 \@ifx{#1\undefined}
}%
\providecommand \@ifnum [1]{%
 \ifnum #1\expandafter \@firstoftwo
 \else \expandafter \@secondoftwo
 \fi
}%
\providecommand \@ifx [1]{%
 \ifx #1\expandafter \@firstoftwo
 \else \expandafter \@secondoftwo
 \fi
}%
\providecommand \natexlab [1]{#1}%
\providecommand \enquote  [1]{``#1''}%
\providecommand \bibnamefont  [1]{#1}%
\providecommand \bibfnamefont [1]{#1}%
\providecommand \citenamefont [1]{#1}%
\providecommand \href@noop [0]{\@secondoftwo}%
\providecommand \href [0]{\begingroup \@sanitize@url \@href}%
\providecommand \@href[1]{\@@startlink{#1}\@@href}%
\providecommand \@@href[1]{\endgroup#1\@@endlink}%
\providecommand \@sanitize@url [0]{\catcode `\\12\catcode `\$12\catcode
  `\&12\catcode `\#12\catcode `\^12\catcode `\_12\catcode `\%12\relax}%
\providecommand \@@startlink[1]{}%
\providecommand \@@endlink[0]{}%
\providecommand \url  [0]{\begingroup\@sanitize@url \@url }%
\providecommand \@url [1]{\endgroup\@href {#1}{\urlprefix }}%
\providecommand \urlprefix  [0]{URL }%
\providecommand \Eprint [0]{\href }%
\providecommand \doibase [0]{http://dx.doi.org/}%
\providecommand \selectlanguage [0]{\@gobble}%
\providecommand \bibinfo  [0]{\@secondoftwo}%
\providecommand \bibfield  [0]{\@secondoftwo}%
\providecommand \translation [1]{[#1]}%
\providecommand \BibitemOpen [0]{}%
\providecommand \bibitemStop [0]{}%
\providecommand \bibitemNoStop [0]{.\EOS\space}%
\providecommand \EOS [0]{\spacefactor3000\relax}%
\providecommand \BibitemShut  [1]{\csname bibitem#1\endcsname}%
\let\auto@bib@innerbib\@empty
\bibitem [{\citenamefont {Lee}\ and\ \citenamefont {Stone}(1985)}]{lee1985}%
  \BibitemOpen
  \bibfield  {author} {\bibinfo {author} {\bibfnamefont {P.~A.}\ \bibnamefont
  {Lee}}\ and\ \bibinfo {author} {\bibfnamefont {A.~D.}\ \bibnamefont
  {Stone}},\ }\bibfield  {title} {\enquote {\bibinfo {title} {Universal
  conductance fluctuations in metals},}\ }\href {\doibase
  10.1103/PhysRevLett.55.1622} {\bibfield  {journal} {\bibinfo  {journal}
  {Phys. Rev. Lett.}\ }\textbf {\bibinfo {volume} {55}},\ \bibinfo {pages}
  {1622--1625} (\bibinfo {year} {1985})}\BibitemShut {NoStop}%
\bibitem [{\citenamefont {Feng}\ \emph {et~al.}(1986)\citenamefont {Feng},
  \citenamefont {Lee},\ and\ \citenamefont {Stone}}]{feng86}%
  \BibitemOpen
  \bibfield  {author} {\bibinfo {author} {\bibfnamefont {S.}~\bibnamefont
  {Feng}}, \bibinfo {author} {\bibfnamefont {P.~A.}\ \bibnamefont {Lee}}, \
  and\ \bibinfo {author} {\bibfnamefont {A.~D.}\ \bibnamefont {Stone}},\
  }\bibfield  {title} {\enquote {\bibinfo {title} {Sensitivity of the
  conductance of a disordered metal to the motion of a single atom:
  Implications for $\frac{1}{f}$ noise},}\ }\href {\doibase
  10.1103/PhysRevLett.56.1960} {\bibfield  {journal} {\bibinfo  {journal}
  {Phys. Rev. Lett.}\ }\textbf {\bibinfo {volume} {56}},\ \bibinfo {pages}
  {1960--1963} (\bibinfo {year} {1986})}\BibitemShut {NoStop}%
\bibitem [{\citenamefont {Jura}\ \emph {et~al.}(2009)\citenamefont {Jura},
  \citenamefont {Topinka}, \citenamefont {Grobis}, \citenamefont {Pfeiffer},
  \citenamefont {West},\ and\ \citenamefont {Goldhaber-Gordon}}]{jura09a}%
  \BibitemOpen
  \bibfield  {author} {\bibinfo {author} {\bibfnamefont {M.~P.}\ \bibnamefont
  {Jura}}, \bibinfo {author} {\bibfnamefont {M.~A.}\ \bibnamefont {Topinka}},
  \bibinfo {author} {\bibfnamefont {M.}~\bibnamefont {Grobis}}, \bibinfo
  {author} {\bibfnamefont {L.~N.}\ \bibnamefont {Pfeiffer}}, \bibinfo {author}
  {\bibfnamefont {K.~W.}\ \bibnamefont {West}}, \ and\ \bibinfo {author}
  {\bibfnamefont {D.}~\bibnamefont {Goldhaber-Gordon}},\ }\bibfield  {title}
  {\enquote {\bibinfo {title} {Electron interferometer formed with a scanning
  probe tip and quantum point contact},}\ }\href {\doibase
  10.1103/PhysRevB.80.041303} {\bibfield  {journal} {\bibinfo  {journal} {Phys.
  Rev. B}\ }\textbf {\bibinfo {volume} {80}},\ \bibinfo {pages} {041303(R)}
  (\bibinfo {year} {2009})}\BibitemShut {NoStop}%
\bibitem [{\citenamefont {Topinka}\ \emph {et~al.}(2000)\citenamefont
  {Topinka}, \citenamefont {LeRoy}, \citenamefont {Shaw}, \citenamefont
  {Heller}, \citenamefont {Westervelt}, \citenamefont {Maranowski},\ and\
  \citenamefont {Gossard}}]{topinka2000science}%
  \BibitemOpen
  \bibfield  {author} {\bibinfo {author} {\bibfnamefont {M.~A.}\ \bibnamefont
  {Topinka}}, \bibinfo {author} {\bibfnamefont {B.~J.}\ \bibnamefont {LeRoy}},
  \bibinfo {author} {\bibfnamefont {S.~E.~J.}\ \bibnamefont {Shaw}}, \bibinfo
  {author} {\bibfnamefont {E.~J.}\ \bibnamefont {Heller}}, \bibinfo {author}
  {\bibfnamefont {R.~M.}\ \bibnamefont {Westervelt}}, \bibinfo {author}
  {\bibfnamefont {K.~D.}\ \bibnamefont {Maranowski}}, \ and\ \bibinfo {author}
  {\bibfnamefont {A.~C.}\ \bibnamefont {Gossard}},\ }\bibfield  {title}
  {\enquote {\bibinfo {title} {Imaging coherent electron flow from a quantum
  point contact},}\ }\href {\doibase 10.1126/science.289.5488.2323} {\bibfield
  {journal} {\bibinfo  {journal} {Science}\ }\textbf {\bibinfo {volume}
  {289}},\ \bibinfo {pages} {2323} (\bibinfo {year} {2000})}\BibitemShut
  {NoStop}%
\bibitem [{\citenamefont {Topinka}\ \emph {et~al.}(2001)\citenamefont
  {Topinka}, \citenamefont {LeRoy}, \citenamefont {Westervelt}, \citenamefont
  {Shaw}, \citenamefont {Fleischmann}, \citenamefont {Heller}, \citenamefont
  {Maranowski},\ and\ \citenamefont {Gossard}}]{topinka2001nature}%
  \BibitemOpen
  \bibfield  {author} {\bibinfo {author} {\bibfnamefont {M.~A.}\ \bibnamefont
  {Topinka}}, \bibinfo {author} {\bibfnamefont {B.~J.}\ \bibnamefont {LeRoy}},
  \bibinfo {author} {\bibfnamefont {R.~M.}\ \bibnamefont {Westervelt}},
  \bibinfo {author} {\bibfnamefont {S.~E.~J.}\ \bibnamefont {Shaw}}, \bibinfo
  {author} {\bibfnamefont {R.}~\bibnamefont {Fleischmann}}, \bibinfo {author}
  {\bibfnamefont {E.~J.}\ \bibnamefont {Heller}}, \bibinfo {author}
  {\bibfnamefont {K.~D.}\ \bibnamefont {Maranowski}}, \ and\ \bibinfo {author}
  {\bibfnamefont {A.~C.}\ \bibnamefont {Gossard}},\ }\bibfield  {title}
  {\enquote {\bibinfo {title} {Coherent branched flow in a two-dimensional
  electron gas},}\ }\href {\doibase 10.1038/35065553} {\bibfield  {journal}
  {\bibinfo  {journal} {Nature}\ }\textbf {\bibinfo {volume} {410}},\ \bibinfo
  {pages} {183} (\bibinfo {year} {2001})}\BibitemShut {NoStop}%
\bibitem [{\citenamefont {Sellier}\ \emph {et~al.}(2011)\citenamefont
  {Sellier}, \citenamefont {Hackens}, \citenamefont {Pala}, \citenamefont
  {Martins}, \citenamefont {Baltazar}, \citenamefont {Wallart}, \citenamefont
  {Desplanque}, \citenamefont {Bayot},\ and\ \citenamefont
  {Huant}}]{sellier2011review}%
  \BibitemOpen
  \bibfield  {author} {\bibinfo {author} {\bibfnamefont {H.}~\bibnamefont
  {Sellier}}, \bibinfo {author} {\bibfnamefont {B.}~\bibnamefont {Hackens}},
  \bibinfo {author} {\bibfnamefont {M.~G.}\ \bibnamefont {Pala}}, \bibinfo
  {author} {\bibfnamefont {F.}~\bibnamefont {Martins}}, \bibinfo {author}
  {\bibfnamefont {S.}~\bibnamefont {Baltazar}}, \bibinfo {author}
  {\bibfnamefont {X.}~\bibnamefont {Wallart}}, \bibinfo {author} {\bibfnamefont
  {L.}~\bibnamefont {Desplanque}}, \bibinfo {author} {\bibfnamefont
  {V.}~\bibnamefont {Bayot}}, \ and\ \bibinfo {author} {\bibfnamefont
  {S.}~\bibnamefont {Huant}},\ }\bibfield  {title} {\enquote {\bibinfo {title}
  {On the imaging of electron transport in semiconductor quantum structures by
  scanning-gate microscopy: successes and limitations},}\ }\href {\doibase
  10.1088/0268-1242/26/6/064008} {\bibfield  {journal} {\bibinfo  {journal}
  {Semicond. Sci. Technol.}\ }\textbf {\bibinfo {volume} {26}},\ \bibinfo
  {pages} {064008} (\bibinfo {year} {2011})}\BibitemShut {NoStop}%
\bibitem [{\citenamefont {Jalabert}\ \emph {et~al.}(2010)\citenamefont
  {Jalabert}, \citenamefont {Szewc}, \citenamefont {Tomsovic},\ and\
  \citenamefont {Weinmann}}]{jalabert2010}%
  \BibitemOpen
  \bibfield  {author} {\bibinfo {author} {\bibfnamefont {R.~A.}\ \bibnamefont
  {Jalabert}}, \bibinfo {author} {\bibfnamefont {W.}~\bibnamefont {Szewc}},
  \bibinfo {author} {\bibfnamefont {S.}~\bibnamefont {Tomsovic}}, \ and\
  \bibinfo {author} {\bibfnamefont {D.}~\bibnamefont {Weinmann}},\ }\bibfield
  {title} {\enquote {\bibinfo {title} {What is measured in the scanning gate
  microscopy of a quantum point contact?}}\ }\href {\doibase
  10.1103/PhysRevLett.105.166802} {\bibfield  {journal} {\bibinfo  {journal}
  {Phys. Rev. Lett.}\ }\textbf {\bibinfo {volume} {105}},\ \bibinfo {pages}
  {166802} (\bibinfo {year} {2010})}\BibitemShut {NoStop}%
\bibitem [{\citenamefont {Gorini}\ \emph {et~al.}(2013)\citenamefont {Gorini},
  \citenamefont {Jalabert}, \citenamefont {Szewc}, \citenamefont {Tomsovic},\
  and\ \citenamefont {Weinmann}}]{gorini2013}%
  \BibitemOpen
  \bibfield  {author} {\bibinfo {author} {\bibfnamefont {C.}~\bibnamefont
  {Gorini}}, \bibinfo {author} {\bibfnamefont {R.~A.}\ \bibnamefont
  {Jalabert}}, \bibinfo {author} {\bibfnamefont {W.}~\bibnamefont {Szewc}},
  \bibinfo {author} {\bibfnamefont {S.}~\bibnamefont {Tomsovic}}, \ and\
  \bibinfo {author} {\bibfnamefont {D.}~\bibnamefont {Weinmann}},\ }\bibfield
  {title} {\enquote {\bibinfo {title} {Theory of scanning gate microscopy},}\
  }\href {\doibase 10.1103/PhysRevB.88.035406} {\bibfield  {journal} {\bibinfo
  {journal} {Phys. Rev. B}\ }\textbf {\bibinfo {volume} {88}},\ \bibinfo
  {pages} {035406} (\bibinfo {year} {2013})}\BibitemShut {NoStop}%
\bibitem [{\citenamefont {Steinacher}\ \emph {et~al.}(2018)\citenamefont
  {Steinacher}, \citenamefont {P{\"o}ltl}, \citenamefont {Kr{\"a}henmann},
  \citenamefont {Hofmann}, \citenamefont {Reichl}, \citenamefont {Zwerger},
  \citenamefont {Wegscheider}, \citenamefont {Jalabert}, \citenamefont
  {Ensslin}, \citenamefont {Weinmann},\ and\ \citenamefont
  {Ihn}}]{steinacher2018}%
  \BibitemOpen
  \bibfield  {author} {\bibinfo {author} {\bibfnamefont {R.}~\bibnamefont
  {Steinacher}}, \bibinfo {author} {\bibfnamefont {C.}~\bibnamefont
  {P{\"o}ltl}}, \bibinfo {author} {\bibfnamefont {T.}~\bibnamefont
  {Kr{\"a}henmann}}, \bibinfo {author} {\bibfnamefont {A.}~\bibnamefont
  {Hofmann}}, \bibinfo {author} {\bibfnamefont {C.}~\bibnamefont {Reichl}},
  \bibinfo {author} {\bibfnamefont {W.}~\bibnamefont {Zwerger}}, \bibinfo
  {author} {\bibfnamefont {W.}~\bibnamefont {Wegscheider}}, \bibinfo {author}
  {\bibfnamefont {R.~A.}\ \bibnamefont {Jalabert}}, \bibinfo {author}
  {\bibfnamefont {K.}~\bibnamefont {Ensslin}}, \bibinfo {author} {\bibfnamefont
  {D.}~\bibnamefont {Weinmann}}, \ and\ \bibinfo {author} {\bibfnamefont
  {T.}~\bibnamefont {Ihn}},\ }\bibfield  {title} {\enquote {\bibinfo {title}
  {Scanning gate experiments: From strongly to weakly invasive probes},}\
  }\href {\doibase 10.1103/PhysRevB.98.075426} {\bibfield  {journal} {\bibinfo
  {journal} {Phys. Rev. B}\ }\textbf {\bibinfo {volume} {98}},\ \bibinfo
  {pages} {075426} (\bibinfo {year} {2018})}\BibitemShut {NoStop}%
\bibitem [{\citenamefont {Braem}\ \emph {et~al.}(2018)\citenamefont {Braem},
  \citenamefont {Gold}, \citenamefont {Hennel}, \citenamefont {R{\"o}{\"o}sli},
  \citenamefont {Berl}, \citenamefont {Dietsche}, \citenamefont {Wegscheider},
  \citenamefont {Ensslin},\ and\ \citenamefont {Ihn}}]{braem2018}%
  \BibitemOpen
  \bibfield  {author} {\bibinfo {author} {\bibfnamefont {B.~A.}\ \bibnamefont
  {Braem}}, \bibinfo {author} {\bibfnamefont {C.}~\bibnamefont {Gold}},
  \bibinfo {author} {\bibfnamefont {S.}~\bibnamefont {Hennel}}, \bibinfo
  {author} {\bibfnamefont {M.}~\bibnamefont {R{\"o}{\"o}sli}}, \bibinfo
  {author} {\bibfnamefont {M.}~\bibnamefont {Berl}}, \bibinfo {author}
  {\bibfnamefont {W.}~\bibnamefont {Dietsche}}, \bibinfo {author}
  {\bibfnamefont {W.}~\bibnamefont {Wegscheider}}, \bibinfo {author}
  {\bibfnamefont {K.}~\bibnamefont {Ensslin}}, \ and\ \bibinfo {author}
  {\bibfnamefont {T.}~\bibnamefont {Ihn}},\ }\bibfield  {title} {\enquote
  {\bibinfo {title} {Stable branched electron flow},}\ }\href {\doibase
  10.1088/1367-2630/aad068} {\bibfield  {journal} {\bibinfo  {journal} {New J.
  Phys.}\ }\textbf {\bibinfo {volume} {20}},\ \bibinfo {pages} {073015}
  (\bibinfo {year} {2018})}\BibitemShut {NoStop}%
\bibitem [{\citenamefont {Fratus}\ \emph {et~al.}(2019)\citenamefont {Fratus},
  \citenamefont {Jalabert},\ and\ \citenamefont {Weinmann}}]{fratus2019}%
  \BibitemOpen
  \bibfield  {author} {\bibinfo {author} {\bibfnamefont {K.~R.}\ \bibnamefont
  {Fratus}}, \bibinfo {author} {\bibfnamefont {R.~A.}\ \bibnamefont
  {Jalabert}}, \ and\ \bibinfo {author} {\bibfnamefont {D.}~\bibnamefont
  {Weinmann}},\ }\bibfield  {title} {\enquote {\bibinfo {title} {Energy
  stability of branching in the scanning gate response of two-dimensional
  electron gases with smooth disorder},}\ }\href {\doibase
  10.1103/PhysRevB.100.155435} {\bibfield  {journal} {\bibinfo  {journal}
  {Phys. Rev. B}\ }\textbf {\bibinfo {volume} {100}},\ \bibinfo {pages}
  {155435} (\bibinfo {year} {2019})}\BibitemShut {NoStop}%
\bibitem [{\citenamefont {Goodfellow}\ \emph {et~al.}(2016)\citenamefont
  {Goodfellow}, \citenamefont {Bengio},\ and\ \citenamefont
  {Courville}}]{deeplearning2016}%
  \BibitemOpen
  \bibfield  {author} {\bibinfo {author} {\bibfnamefont {I.}~\bibnamefont
  {Goodfellow}}, \bibinfo {author} {\bibfnamefont {Y.}~\bibnamefont {Bengio}},
  \ and\ \bibinfo {author} {\bibfnamefont {A.}~\bibnamefont {Courville}},\
  }\href@noop {} {\emph {\bibinfo {title} {Deep Learning}}}\ (\bibinfo
  {publisher} {MIT Press},\ \bibinfo {year} {2016})\ \bibinfo {note}
  {\url{http://www.deeplearningbook.org}}\BibitemShut {NoStop}%
\bibitem [{\citenamefont {McCulloch}\ and\ \citenamefont
  {Pitts}(1943)}]{mcculloch1943}%
  \BibitemOpen
  \bibfield  {author} {\bibinfo {author} {\bibfnamefont {W.~S.}\ \bibnamefont
  {McCulloch}}\ and\ \bibinfo {author} {\bibfnamefont {W.}~\bibnamefont
  {Pitts}},\ }\bibfield  {title} {\enquote {\bibinfo {title} {A logical
  calculus of the ideas immanent in nervous activity},}\ }\href {\doibase
  10.1007/BF02478259} {\bibfield  {journal} {\bibinfo  {journal} {The bulletin
  of mathematical biophysics}\ }\textbf {\bibinfo {volume} {5}},\ \bibinfo
  {pages} {115} (\bibinfo {year} {1943})}\BibitemShut {NoStop}%
\bibitem [{\citenamefont {Carleo}\ \emph {et~al.}(2019)\citenamefont {Carleo},
  \citenamefont {Cirac}, \citenamefont {Cranmer}, \citenamefont {Daudet},
  \citenamefont {Schuld}, \citenamefont {Tishby}, \citenamefont
  {Vogt-Maranto},\ and\ \citenamefont {Zdeborov\'a}}]{ML-RevModPhys2019}%
  \BibitemOpen
  \bibfield  {author} {\bibinfo {author} {\bibfnamefont {G.}~\bibnamefont
  {Carleo}}, \bibinfo {author} {\bibfnamefont {I.}~\bibnamefont {Cirac}},
  \bibinfo {author} {\bibfnamefont {K.}~\bibnamefont {Cranmer}}, \bibinfo
  {author} {\bibfnamefont {L.}~\bibnamefont {Daudet}}, \bibinfo {author}
  {\bibfnamefont {M.}~\bibnamefont {Schuld}}, \bibinfo {author} {\bibfnamefont
  {N.}~\bibnamefont {Tishby}}, \bibinfo {author} {\bibfnamefont
  {L.}~\bibnamefont {Vogt-Maranto}}, \ and\ \bibinfo {author} {\bibfnamefont
  {L.}~\bibnamefont {Zdeborov\'a}},\ }\bibfield  {title} {\enquote {\bibinfo
  {title} {Machine learning and the physical sciences},}\ }\href {\doibase
  10.1103/RevModPhys.91.045002} {\bibfield  {journal} {\bibinfo  {journal}
  {Rev. Mod. Phys.}\ }\textbf {\bibinfo {volume} {91}},\ \bibinfo {pages}
  {045002} (\bibinfo {year} {2019})}\BibitemShut {NoStop}%
\bibitem [{\citenamefont {Iten}\ \emph {et~al.}(2020)\citenamefont {Iten},
  \citenamefont {Metger}, \citenamefont {Wilming}, \citenamefont {del Rio},\
  and\ \citenamefont {Renner}}]{ML-concepts2020}%
  \BibitemOpen
  \bibfield  {author} {\bibinfo {author} {\bibfnamefont {R.}~\bibnamefont
  {Iten}}, \bibinfo {author} {\bibfnamefont {T.}~\bibnamefont {Metger}},
  \bibinfo {author} {\bibfnamefont {H.}~\bibnamefont {Wilming}}, \bibinfo
  {author} {\bibfnamefont {L.}~\bibnamefont {del Rio}}, \ and\ \bibinfo
  {author} {\bibfnamefont {R.}~\bibnamefont {Renner}},\ }\bibfield  {title}
  {\enquote {\bibinfo {title} {Discovering physical concepts with neural
  networks},}\ }\href {\doibase 10.1103/PhysRevLett.124.010508} {\bibfield
  {journal} {\bibinfo  {journal} {Phys. Rev. Lett.}\ }\textbf {\bibinfo
  {volume} {124}},\ \bibinfo {pages} {010508} (\bibinfo {year}
  {2020})}\BibitemShut {NoStop}%
\bibitem [{\citenamefont {Voosen}(2017)}]{voosen_XAI2017}%
  \BibitemOpen
  \bibfield  {author} {\bibinfo {author} {\bibfnamefont {P.}~\bibnamefont
  {Voosen}},\ }\bibfield  {title} {\enquote {\bibinfo {title} {The {AI}
  detectives},}\ }\href {\doibase 10.1126/science.357.6346.22} {\bibfield
  {journal} {\bibinfo  {journal} {Science}\ }\textbf {\bibinfo {volume}
  {357}},\ \bibinfo {pages} {22--27} (\bibinfo {year} {2017})}\BibitemShut
  {NoStop}%
\bibitem [{\citenamefont {Patsyk}\ \emph {et~al.}(2020)\citenamefont {Patsyk},
  \citenamefont {Sivan}, \citenamefont {Segev},\ and\ \citenamefont
  {Bandres}}]{patsyk20}%
  \BibitemOpen
  \bibfield  {author} {\bibinfo {author} {\bibfnamefont {A.}~\bibnamefont
  {Patsyk}}, \bibinfo {author} {\bibfnamefont {U.}~\bibnamefont {Sivan}},
  \bibinfo {author} {\bibfnamefont {M.}~\bibnamefont {Segev}}, \ and\ \bibinfo
  {author} {\bibfnamefont {M.~A.}\ \bibnamefont {Bandres}},\ }\bibfield
  {title} {\enquote {\bibinfo {title} {Observation of branched flow of
  light},}\ }\href {\doibase 10.1038/s41586-020-2376-8} {\bibfield  {journal}
  {\bibinfo  {journal} {Nature}\ }\textbf {\bibinfo {volume} {583}},\ \bibinfo
  {pages} {60} (\bibinfo {year} {2020})}\BibitemShut {NoStop}%
\bibitem [{\citenamefont {Degueldre}\ \emph {et~al.}(2016)\citenamefont
  {Degueldre}, \citenamefont {Metzger}, \citenamefont {Geisel},\ and\
  \citenamefont {Fleischmann}}]{degueldre_tsunami2016}%
  \BibitemOpen
  \bibfield  {author} {\bibinfo {author} {\bibfnamefont {H.}~\bibnamefont
  {Degueldre}}, \bibinfo {author} {\bibfnamefont {J.~J.}\ \bibnamefont
  {Metzger}}, \bibinfo {author} {\bibfnamefont {T.}~\bibnamefont {Geisel}}, \
  and\ \bibinfo {author} {\bibfnamefont {R.}~\bibnamefont {Fleischmann}},\
  }\bibfield  {title} {\enquote {\bibinfo {title} {Random focusing of tsunami
  waves},}\ }\href {\doibase 10.1038/nphys3557} {\bibfield  {journal} {\bibinfo
   {journal} {Nature Physics}\ }\textbf {\bibinfo {volume} {12}},\ \bibinfo
  {pages} {259} (\bibinfo {year} {2016})}\BibitemShut {NoStop}%
\bibitem [{\citenamefont {Groth}\ \emph {et~al.}(2014)\citenamefont {Groth},
  \citenamefont {Wimmer}, \citenamefont {Akhmerov},\ and\ \citenamefont
  {Waintal}}]{KWANT_2014}%
  \BibitemOpen
  \bibfield  {author} {\bibinfo {author} {\bibfnamefont {C.~W.}\ \bibnamefont
  {Groth}}, \bibinfo {author} {\bibfnamefont {M.}~\bibnamefont {Wimmer}},
  \bibinfo {author} {\bibfnamefont {A.~R.}\ \bibnamefont {Akhmerov}}, \ and\
  \bibinfo {author} {\bibfnamefont {X.}~\bibnamefont {Waintal}},\ }\bibfield
  {title} {\enquote {\bibinfo {title} {Kwant: a software package for quantum
  transport},}\ }\href {\doibase 10.1088/1367-2630/16/6/063065} {\bibfield
  {journal} {\bibinfo  {journal} {New J. Phys.}\ }\textbf {\bibinfo {volume}
  {16}},\ \bibinfo {pages} {063065} (\bibinfo {year} {2014})}\BibitemShut
  {NoStop}%
\bibitem [{\citenamefont {Ihn}(2010)}]{ihn2010semiconductor}%
  \BibitemOpen
  \bibfield  {author} {\bibinfo {author} {\bibfnamefont {T.}~\bibnamefont
  {Ihn}},\ }\href {\doibase 10.1093/acprof:oso/9780199534425.001.0001} {\emph
  {\bibinfo {title} {Semiconductor Nanostructures: Quantum states and
  electronic transport}}}\ (\bibinfo  {publisher} {Oxford University Press},\
  \bibinfo {year} {2010})\BibitemShut {NoStop}%
\bibitem [{\citenamefont {B{\"u}ttiker}\ and\ \citenamefont
  {Christen}(1996)}]{buettiker1996}%
  \BibitemOpen
  \bibfield  {author} {\bibinfo {author} {\bibfnamefont {M.}~\bibnamefont
  {B{\"u}ttiker}}\ and\ \bibinfo {author} {\bibfnamefont {T.}~\bibnamefont
  {Christen}},\ }\enquote {\bibinfo {title} {Basic elements of electrical
  conduction},}\ in\ \href {\doibase 10.1007/978-94-009-1760-6_13} {\emph
  {\bibinfo {booktitle} {Quantum Transport in Semiconductor Submicron
  Structures}}},\ \bibinfo {editor} {edited by\ \bibinfo {editor}
  {\bibfnamefont {B.}~\bibnamefont {Kramer}}}\ (\bibinfo  {publisher} {Springer
  Netherlands},\ \bibinfo {address} {Dordrecht},\ \bibinfo {year} {1996})\ pp.\
  \bibinfo {pages} {263--291}\BibitemShut {NoStop}%
\bibitem [{\citenamefont {Gramespacher}\ and\ \citenamefont
  {B\"uttiker}(1999)}]{gramespacher1999}%
  \BibitemOpen
  \bibfield  {author} {\bibinfo {author} {\bibfnamefont {T.}~\bibnamefont
  {Gramespacher}}\ and\ \bibinfo {author} {\bibfnamefont {M.}~\bibnamefont
  {B\"uttiker}},\ }\bibfield  {title} {\enquote {\bibinfo {title} {Local
  densities, distribution functions, and wave-function correlations for
  spatially resolved shot noise at nanocontacts},}\ }\href {\doibase
  10.1103/PhysRevB.60.2375} {\bibfield  {journal} {\bibinfo  {journal} {Phys.
  Rev. B}\ }\textbf {\bibinfo {volume} {60}},\ \bibinfo {pages} {2375}
  (\bibinfo {year} {1999})}\BibitemShut {NoStop}%
\bibitem [{\citenamefont {Ly}\ \emph {et~al.}(2017)\citenamefont {Ly},
  \citenamefont {Jalabert}, \citenamefont {Tomsovic},\ and\ \citenamefont
  {Weinmann}}]{Ly2017}%
  \BibitemOpen
  \bibfield  {author} {\bibinfo {author} {\bibfnamefont {O.}~\bibnamefont
  {Ly}}, \bibinfo {author} {\bibfnamefont {R.~A.}\ \bibnamefont {Jalabert}},
  \bibinfo {author} {\bibfnamefont {S.}~\bibnamefont {Tomsovic}}, \ and\
  \bibinfo {author} {\bibfnamefont {D.}~\bibnamefont {Weinmann}},\ }\bibfield
  {title} {\enquote {\bibinfo {title} {Partial local density of states from
  scanning gate microscopy},}\ }\href {\doibase 10.1103/PhysRevB.96.125439}
  {\bibfield  {journal} {\bibinfo  {journal} {Phys. Rev. B}\ }\textbf {\bibinfo
  {volume} {96}},\ \bibinfo {pages} {125439} (\bibinfo {year}
  {2017})}\BibitemShut {NoStop}%
\bibitem [{\citenamefont {Fratus}\ \emph {et~al.}(2021)\citenamefont {Fratus},
  \citenamefont {Calonnec}, \citenamefont {Jalabert}, \citenamefont {Weick},\
  and\ \citenamefont {Weinmann}}]{Keith_2021}%
  \BibitemOpen
  \bibfield  {author} {\bibinfo {author} {\bibfnamefont {K.~R.}\ \bibnamefont
  {Fratus}}, \bibinfo {author} {\bibfnamefont {C.~Le}\ \bibnamefont
  {Calonnec}}, \bibinfo {author} {\bibfnamefont {R.~A.}\ \bibnamefont
  {Jalabert}}, \bibinfo {author} {\bibfnamefont {G.}~\bibnamefont {Weick}}, \
  and\ \bibinfo {author} {\bibfnamefont {D.}~\bibnamefont {Weinmann}},\
  }\bibfield  {title} {\enquote {\bibinfo {title} {{Signatures of folded
  branches in the scanning gate microscopy of ballistic electronic
  cavities}},}\ }\href {\doibase 10.21468/SciPostPhys.10.3.069} {\bibfield
  {journal} {\bibinfo  {journal} {SciPost Phys.}\ }\textbf {\bibinfo {volume}
  {10}},\ \bibinfo {pages} {69} (\bibinfo {year} {2021})}\BibitemShut {NoStop}%
\bibitem [{\citenamefont {LeCun}\ \emph {et~al.}(2015)\citenamefont {LeCun},
  \citenamefont {Bengio},\ and\ \citenamefont {Hinton}}]{LeCun2015}%
  \BibitemOpen
  \bibfield  {author} {\bibinfo {author} {\bibfnamefont {Y.}~\bibnamefont
  {LeCun}}, \bibinfo {author} {\bibfnamefont {Y.}~\bibnamefont {Bengio}}, \
  and\ \bibinfo {author} {\bibfnamefont {G.}~\bibnamefont {Hinton}},\
  }\bibfield  {title} {\enquote {\bibinfo {title} {Deep learning},}\ }\href
  {\doibase 10.1038/nature14539} {\bibfield  {journal} {\bibinfo  {journal}
  {Nature}\ }\textbf {\bibinfo {volume} {521}},\ \bibinfo {pages} {436--444}
  (\bibinfo {year} {2015})}\BibitemShut {NoStop}%
\bibitem [{Note1()}]{Note1}%
  \BibitemOpen
  \bibinfo {note} {In this work we use an image which corresponds to a 2D
  array, but CNN can also handle 3D arrays.}\BibitemShut {Stop}%
\bibitem [{Note2()}]{Note2}%
  \BibitemOpen
  \bibinfo {note} {The test set is an ensemble of samples similar to the ones
  which compose the training set. However, they are not used in the training
  process.}\BibitemShut {Stop}%
\bibitem [{Note3()}]{Note3}%
  \BibitemOpen
  \bibinfo {note} {This kind of procedure using neural networks and physical
  equations ressembles to a physics-informed neural network which combines
  neural networks and non-linear differential equations to give a physically
  correct result \cite {Karniadakis2021}}\BibitemShut {NoStop}%
\bibitem [{\citenamefont {Percebois}\ \emph {et~al.}()\citenamefont {Percebois}
  \emph {et~al.}}]{percebois_in_progress}%
  \BibitemOpen
  \bibfield  {author} {\bibinfo {author} {\bibfnamefont {G.~J.}\ \bibnamefont
  {Percebois}} \emph {et~al.},\ }\href@noop {} {}\bibinfo {note} {{w}ork in
  progress (2021)}\BibitemShut {NoStop}%
\bibitem [{Note4()}]{Note4}%
  \BibitemOpen
  \bibinfo {note} {The discrepancy is evaluated quantitatively through a loss
  function. For the regression problem of the present section, the loss
  function we use is the squared error, averaged over the elements of the
  training set.}\BibitemShut {Stop}%
\bibitem [{Note5()}]{Note5}%
  \BibitemOpen
  \bibinfo {note} {The convolutional layer applies filters on the previous
  images in order to detect particular features in the images. The main
  characteristics of a convolution layer is the size of the filters and the
  number of filters.}\BibitemShut {Stop}%
\bibitem [{Note6()}]{Note6}%
  \BibitemOpen
  \bibinfo {note} {The aim of the pooling layer is to reduce the dimension
  (i.e., the height and the width) of the previous layer in order to keep only
  the relevant information. This reduction condenses the values of a small
  cluster of pixels of the previous layer in a single value. A pooling layer is
  characterized by the number of pixels which compose the cluster and the
  function applied on the cluster to reduce it to a single number. In our study
  we always use the MaxPooling which consists in taking the largest value of
  the cluster.}\BibitemShut {Stop}%
\bibitem [{\citenamefont {Aggarwal}(2018)}]{aggarwal2018}%
  \BibitemOpen
  \bibfield  {author} {\bibinfo {author} {\bibfnamefont {C.~C.}\ \bibnamefont
  {Aggarwal}},\ }\href
  {https://link.springer.com/book/10.1007/978-3-319-94463-0} {\emph {\bibinfo
  {title} {Neural Networks and Deep Learning}}}\ (\bibinfo  {publisher}
  {Springer},\ \bibinfo {year} {2018})\ p.\ \bibinfo {pages} {127}\BibitemShut
  {NoStop}%
\bibitem [{\citenamefont {Kingma}\ and\ \citenamefont {Ba}()}]{Adam_2014}%
  \BibitemOpen
  \bibfield  {author} {\bibinfo {author} {\bibfnamefont {D.~P.}\ \bibnamefont
  {Kingma}}\ and\ \bibinfo {author} {\bibfnamefont {J.}~\bibnamefont {Ba}},\
  }\href {https://arxiv.org/abs/1412.6980} {\enquote {\bibinfo {title} {Adam: A
  method for stochastic optimization},}\ }\Eprint
  {http://arxiv.org/abs/1412.6980} {arXiv:1412.6980} \BibitemShut {NoStop}%
\bibitem [{Note7()}]{Note7}%
  \BibitemOpen
  \bibinfo {note} {The number of epochs corresponds to the number of times that
  the entire training set is used to adjust the parameters of the neural
  network.}\BibitemShut {Stop}%
\bibitem [{\citenamefont {Chollet}\ \emph {et~al.}(2015)\citenamefont {Chollet}
  \emph {et~al.}}]{chollet2015keras}%
  \BibitemOpen
  \bibfield  {author} {\bibinfo {author} {\bibfnamefont {F.}~\bibnamefont
  {Chollet}} \emph {et~al.},\ }\href {https://keras.io} {\enquote {\bibinfo
  {title} {Keras},}\ } (\bibinfo {year} {2015})\BibitemShut {NoStop}%
\bibitem [{Note8()}]{Note8}%
  \BibitemOpen
  \bibinfo {note} {The UpSampling layer is the inverse of a pooling
  layer.}\BibitemShut {Stop}%
\bibitem [{\citenamefont {Karniadakis}\ \emph {et~al.}(2021)\citenamefont
  {Karniadakis}, \citenamefont {Kevrekidis}, \citenamefont {Lu}, \citenamefont
  {Perdikaris}, \citenamefont {Wang},\ and\ \citenamefont
  {Yang}}]{Karniadakis2021}%
  \BibitemOpen
  \bibfield  {author} {\bibinfo {author} {\bibfnamefont {G.~E.}\ \bibnamefont
  {Karniadakis}}, \bibinfo {author} {\bibfnamefont {I.~G.}\ \bibnamefont
  {Kevrekidis}}, \bibinfo {author} {\bibfnamefont {L.}~\bibnamefont {Lu}},
  \bibinfo {author} {\bibfnamefont {P.}~\bibnamefont {Perdikaris}}, \bibinfo
  {author} {\bibfnamefont {S.}~\bibnamefont {Wang}}, \ and\ \bibinfo {author}
  {\bibfnamefont {L.}~\bibnamefont {Yang}},\ }\bibfield  {title} {\enquote
  {\bibinfo {title} {Physics-informed machine learning},}\ }\href {\doibase
  10.1038/s42254-021-00314-5} {\bibfield  {journal} {\bibinfo  {journal} {Nat.
  Rev. Phys.}\ }\textbf {\bibinfo {volume} {1322}},\ \bibinfo {pages}
  {422--440} (\bibinfo {year} {2021})}\BibitemShut {NoStop}%
\end{thebibliography}%

\end{document}